\documentclass[pre, twocolumn,showkeys,groupaddress,preprintnumbers,floatfix,superscriptaddress]{revtex4-1}

\usepackage{etex}                        \usepackage{ifpdf}                       \usepackage{xspace}                      \usepackage[table]{xcolor}               \usepackage{setspace}                    
\PassOptionsToPackage{final}{graphicx}   \usepackage{graphicx}                    
\makeatletter                            \@ifclassloaded{beamer}{}{\PassOptionsToPackage{pagebackref}{hyperref}}
\makeatother
\usepackage{hyperref}
\usepackage{url}                         \usepackage[super]{nth}                  \usepackage{contour}                     \usepackage{verbatim}                    \usepackage{ragged2e}                    \usepackage{attrib}                      \usepackage{adjustbox}                   \usepackage[absolute,overlay]{textpos}   \usepackage{calc}                        
\usepackage{amsmath}                     \usepackage{amscd}                       \usepackage{amsfonts}                    \usepackage{amssymb}                     \usepackage{amsthm}                      \usepackage{scalerel}                    \usepackage{bm}                          \usepackage{bbm}                         \usepackage{braket}				               \usepackage{mathtools}                   \usepackage{etoolbox}                    \usepackage{textcomp}                    \usepackage{stmaryrd}                    \usepackage{nicefrac}                    \usepackage[binary-units]{siunitx}       
\usepackage[T1]{fontenc}                 \usepackage{lmodern}                     \usepackage[scaled=0.72]{beramono}       \usepackage{microtype}                   
\usepackage{algorithm}                   \usepackage{algorithmicx}                
\usepackage{booktabs}                    \usepackage{dcolumn}                     
\makeatletter
  \@ifclassloaded{revtex4-1}
  {
    \usepackage[caption=false]{subfig}                                              }
  {
    \usepackage{caption}
    \usepackage{subcaption}
  }
  \@ifclassloaded{beamer}
  {
    \usepackage[backend=biber,
                autocite=superscript,
                doi=false,
                url=false,
                sorting=none]{biblatex}
    \usepackage[useregional,showdow]{datetime2}
  }
  {
        \usepackage{enumitem}                  }
\makeatother

\definecolor{ryan}{RGB}{64, 0, 64}
\definecolor{nix}{RGB}{255, 0, 0}

\definecolor{ucdblue1}{cmyk}{.87,.46,0,.49} \definecolor{ucdblue2}{cmyk}{1., .56, 0., .34}
\colorlet{ucdblue}{ucdblue2}

\makeatletter
\def\@lox@prtc{\section*{\@fxlistfixmename}\begingroup\def\@dotsep{4.5}}
\def\@lox@psttc{\endgroup}
\makeatother

\usepackage{epigraph}

\colorlet {past_color}    {red}
\colorlet {pres_color}    {blue}
\colorlet {futu_color}    {black!30!green}

\colorlet {temp_color_1}  {red!50!blue}
\colorlet {temp_color_2}  {red!50!green}
\colorlet {temp_color_3}  {blue!50!green}

\colorlet {hmu_color}     {blue!67!green}
\colorlet {rhomu_color}   {temp_color_1!80!blue}
\colorlet {rmu_color}     {blue}
\colorlet {bmu_1_color}   {temp_color_1}
\colorlet {bmu_2_color}   {temp_color_3}
\colorlet {qmu_color}     {temp_color_1!67!green}
\colorlet {wmu_color}     {temp_color_2!57!blue}
\colorlet {sigmamu_color} {temp_color_2}

\usepackage{listings}
\lstdefinestyle{mypython}{
language=Python,                        basicstyle=\small\ttfamily,             keywordstyle=\color{green!50!black},    commentstyle=\color{gray},              numbers=left,                           numberstyle=\tiny,                      stepnumber=1,                           numbersep=5pt,                          backgroundcolor=\color{gray!10},        frame=none,                             tabsize=2,                              captionpos=b,                           breaklines=true,                        breakatwhitespace=false,                showspaces=false,                       showtabs=false,                         morekeywords={as},                      }

\theoremstyle{plain}    \newtheorem{Lem}{Lemma}
\theoremstyle{plain}    
\theoremstyle{plain}    \newtheorem{Cor}{Corollary}
\theoremstyle{plain}    
\theoremstyle{plain}    \newtheorem{The}{Theorem}
\theoremstyle{plain}    \newtheorem*{ProThe}{Proof}
\theoremstyle{plain}    
\theoremstyle{plain}    
\theoremstyle{plain}    
\theoremstyle{plain}    
\theoremstyle{plain}    
\theoremstyle{plain}

\newcommand{\CausalState}       { \mathcal{S} }

\newcommand{\forward}{+}
\newcommand{\reverse}{-}
\newcommand{\forwardreverse}{\pm} 

\newcommand{\FutureCausalState} { {\CausalState}^{\forward} }

\newcommand{\PastCausalState}   { {\CausalState}^{\reverse} }

\newcommand{\lastindex}[2]{
  \edef\tempa{0}
  \edef\tempb{#2}
  \ifx\tempa\tempb
        \edef\tempc{#1}
  \else
        \edef\tempa{0}
    \edef\tempb{#1}
    \ifx\tempa\tempb
      \edef\tempc{#2}
    \else
      \edef\tempc{#1+#2}
    \fi
  \fi
  \tempc
}

\newcommand{\CSjoint}[1][,]{
   \edef\tempa{:}
   \edef\tempb{#1}
   \ifx\tempa\tempb
            \ensuremath{\FutureCausalState\!#1\PastCausalState}
   \else
            \ensuremath{\FutureCausalState#1\PastCausalState}
   \fi
}
\newcommand{\CSjointKL}[3][,]{
   \edef\tempa{:}
   \edef\tempb{#1}
   \ifx\tempa\tempb
            \ensuremath{\FutureCausalState_{#2}\!#1\PastCausalState_{#3}}
   \else
            \ensuremath{\FutureCausalState_{#2}#1\PastCausalState_{#3}}
   \fi
}

\newif\ifpm
\edef\tempa{\forwardreverse}
\edef\tempb{\pm}
\ifx\tempa\tempb
   \pmfalse
\else
   \pmtrue
\fi

\ifcsname mathclap\endcsname
  \relax
\else
  \def\clap#1{\hbox to 0pt{\hss#1\hss}}

\fi

\newcommand{\op} [3] [] {
  \ensuremath{
    \operatorname{#2_{#1}}
    \if\relax\detokenize{#3}\relax
    \else
      \left[ #3 \right]
    \fi
  }
  \xspace
}

\makeatletter
  \@ifclassloaded{beamer}
  {
    \hypersetup{
      linkcolor={blue!50!black},
      citecolor={blue!50!black},
      urlcolor={blue!80!black},
    }
  }
  {
    \hypersetup{
      colorlinks,
      linkcolor={blue!50!black},
      citecolor={blue!50!black},
      urlcolor={blue!80!black},
    }
  }
\makeatother

\usepackage{empheq}
\usepackage{overpic}
\usepackage{qcircuit}
\usepackage{xcolor}

\newcommand{\argmin}{\text{argmin}}

\newcommand{\kB}{k_\text{B}}
\newcommand{\Wdiss}{W_\text{diss}}

\newcommand{\drive}{x_{0:\tau}}

\newcommand{\stationary}{\boldsymbol{\pi}}

\newcommand{\EP}[1][\rho_0]{\boldsymbol{\Sigma}_{#1}}
\newcommand{\EF}{\boldsymbol{\Phi}}  \newcommand{\tr}{\text{tr}}

\newcommand{\trsys}{\tr_\text{sys}}
\newcommand{\trallbutb}{\tr_{\text{sys}, \mathbb{B} \setminus b}}
\newcommand{\trenv}{\tr_\text{env}}

\newcommand{\grad}{\boldsymbol{\nabla}}

\newcommand{\QProcess}{\Gamma}
\newcommand{\q}{\sigma}

\begin{document}

\def\ourTitle{Initial-State Dependence of Thermodynamic Dissipation for any Quantum Process
}

\def\ourAbstract{New exact results
about the nonequilibrium thermodynamics of open quantum systems at arbitrary timescales
are obtained
by considering all possible variations of initial conditions of a system, its environment, and correlations between them.
First we obtain a new quantum-information theoretic equality for entropy production,
valid for an arbitrary initial joint state of system and environment.
For any finite-time process
with a fixed initial environment, we then show that 
the system's loss of
distinction---relative to the minimally dissipative state---exactly quantifies
its thermodynamic dissipation.
The quantum component of this dissipation is the change in coherence
relative to the minimally dissipative state.
Implications for
quantum state preparation and
local control
are explored.
For nonunitary
processes---like the preparation of any particular quantum state---we find that
mismatched expectations lead to divergent dissipation
as the actual initial state becomes orthogonal to the anticipated one.
}

\def\ourKeywords{  nonequilibrium thermodynamics, entropy
  production, relative entropy, open quantum systems
}

\hypersetup{
  pdfauthor={Paul M. Riechers},
  pdftitle={\ourTitle},
  pdfsubject={\ourAbstract},
  pdfkeywords={\ourKeywords},
  pdfproducer={},
  pdfcreator={}
}

\title{\ourTitle}

\author{Paul M. Riechers}
\email{pmriechers@gmail.com}

\affiliation{School of Physical and Mathematical Sciences, Nanyang Technological University,
637371 Singapore}

\affiliation{Complexity Institute, Nanyang Technological University,
637335 Singapore}

\author{Mile Gu}
\email{ceptryn@gmail.com}

\affiliation{School of Physical and Mathematical Sciences, Nanyang Technological University,
637371 Singapore}

\affiliation{Complexity Institute, Nanyang Technological University,
637335 Singapore}

\affiliation{Centre for Quantum Technologies, National University of Singapore,
3 Science Drive 2,
117543 Singapore}

\date{\today}
\bibliographystyle{unsrt}

\begin{abstract}
\ourAbstract
\end{abstract}


\date{\today}
\maketitle

\setstretch{1.1}

\section{Introduction}

Much recent progress extends
Landauer's principle
to the quantum regime---affirming that quantum information is physical~\cite{Zure03, Del11, Fais15, Parr15a, Gool16}.
Associated bounds
refine our understanding of
how much heat
needs to be exhausted---or how much work needs to be performed, or could be extracted---to preserve the
Second Law of Thermodynamics:
Entropy production is expected to be non-negative  $\EP \geq 0$
from any initial density matrix $\rho_0$.
However, these Landauer-type bounds
only become tight in the infinite-time quasistatic limit,
as entropy production
goes to zero.
Yet infinite time is not a luxury afforded to quantum systems
with short decoherence time.
And, even if coherence can be maintained for significant time-length,
we want to know the thermodynamic limits of both
quantum computers and natural quantum processes that transform quickly.

Here, we demonstrate a source of heat dissipation beyond Landaur's bound that applies at any timescale. 
We illustrate that when engineering any nonunitary process, entropy production always has initial state
dependence. It is impossible to optimize resulting entropy production for all input states. Instead, any choice
of realization implies some minimally dissipative state $\q_0$. The injection of any other input state, $\rho_0$,
results in extra dissipation quantified by 
\begin{align}
\EP - \EP[\q_0] = \kB \text{D} \bigl[ \rho_0 \big\| \q_0 \bigr] - \kB \text{D} \bigl[ \rho_\tau \big\| \q_\tau \bigr] ~,
\label{eq:MainResult1}
\end{align}
the contraction of the relative entropy between the actual input state $\rho_0$
and the minimally dissipative input $\q_0$ over the time-interval $\tau$ in which the process is applied. This
dissipation is additional to that given by Landauer, and generalizes a
theorem by Kolchinsky and Wolpert for classical computation to scenarios where quantum coherence can play a significant role~\cite{Kolc17}.

We then highlight some immediate consequences. First, our result implies
entropy production 
for
almost all inputs to any reset protocol.
Second, it implies a
thermodynamic
cost to misaligned expectations:
To minimize heat dissipation, one should
tailor the implementation of a desired quantum operation to the expected initial-state distribution;
but the same optimization
can lead
to
divergent dissipation when input states differ significantly from predictions.
Third,
we find the general thermodynamic cost of modularity:
quantum
gates optimized for thermal efficiency individually can result in unavoidable
entropy production when placed within a larger quantum circuit.
All of these results are valid over arbitrarily short timescales.

Our approach involves developing a framework to determine how initial conditions of system and environment affect entropy production in general for any finite-time quantum process.
This involves an information-theoretic decomposition of entropy production
(see Eq.~\eqref{eq:EPasCorrPlusNAFE} below) which shows that
entropy production is the change in total correlation among system and baths plus the changes in nonequilibrium addition to free energy of each thermodynamic bath. The framework describes heat and entropy flow in all cases, including those with multiple thermal baths initially out of local equilibrium, correlated with each other and the system of interest. Our results complement and extend
the short list of
exact general results known about the finite-time
nonequilibrium thermodynamics of open
quantum systems~\footnote{More often we must
	rely on approximations---by assuming weak coupling and Markovian dynamics~\cite{Lind76, Alic07}, or leveraging linear response and
	local equilibrium theories~\cite{deGr84}---which have provided practical successes in their domain of applicability~\cite{Kubo66, Zwan65, Alic18},
	but cannot be trusted far from equilibrium.},
including
fluctuation relations~\cite{Croo99a, Talk07, Jarz97a, Tasa00, Parr09, Mori11, Deff11, Aber18, Kwon19, Bind18a, Mica20},
a previous information-theoretic decomposition of entropy production~\cite{Espo10a, Reeb14},
and single-shot results that can be derived from these~\cite{Dahl17, Halp18}.
Collectively,
these nonequilibrium \emph{equalities}
subsume the \emph{inequality} of the Second Law of Thermodynamics,
and guide the understanding of far-from-equilibrium phenomena.

\section{Setup}

We consider
a system's transformation while
a set of time-dependent control parameters $x_t$ changes its Hamiltonian and its
interaction with the environment.
The control protocol $\drive$
induces a net unitary time evolution
$\mathcal{U}_{\drive}$ of the system--environment mega-system,
so that the joint state at the end of the transformation 
is~\footnote{While we only utilize the \emph{existence} of the net unitary time evolution,
we note that it is induced
through the time-ordered exponential
involving the total Hamiltonian
$H_{x_t}^\text{tot}$. 
}
\begin{align}
\rho_{\tau}^\text{tot} = \mathcal{U}_{\drive} \rho_0^\text{tot}
\mathcal{U}_{\drive}^\dagger ~.
\label{eq:NetUnitaryEvolution1}
\end{align}
We will also consider the initial ($t=0$) and final ($t=\tau$) reduced states of the system
$\rho_t =  \trenv ( \rho_t^\text{tot} )$
and environment
$\rho_t^{\text{env}} = \trsys ( \rho_t^\text{tot} ) $.

In thermodynamics,
the expected 
\emph{entropy production} for any process
is given by~\cite{deGr84, Kond14, Espo10a, Deff11} \begin{align}
\EP \equiv \EF + \Delta S(\rho_t) ~,
\label{eq:EPdef}
\end{align}
as the \emph{expected entropy flow} $\EF$ to the environment
plus any change in thermodynamic entropy of the system
$\Delta S(\rho_t) = \kB \tr(\rho_0 \ln \rho_0) - \kB \tr(\rho_\tau \ln \rho_\tau)$,
where $\kB$ is Boltzmann's constant. We find that the expected entropy flow can generally be
represented as
\begin{align}
\EF \equiv \kB \tr(\rho_0^\text{env} \ln \stationary^\text{env}) -  \kB \tr(\rho_\tau^\text{env} \ln \stationary^\text{env}) ~,
\label{eq:GenEFdef}
\end{align}
where $\stationary^\text{env}$ is a
reference state
that represents the environment as a set of thermodynamic baths
$\mathbb{B}$ in local equilibrium:
$\stationary^\text{env} = \bigotimes_{b \in \mathbb{B}} \stationary^{(b)}$.
The equilibrium state $\stationary^{(b)}$
is constructed
with the bath's operators 
(e.g., Hamiltonian $H^{(b)}$, number operators $N^{(b, \ell)}$, etc.) 
that correspond to its variable observable quantities (energy, particle numbers, etc.)~\cite{Reic09, Albe01}.
The initial temperature $T^{(b)}$,
chemical potentials $\{ \mu^{(b, \ell)} \}_\ell$,
etc., are fixed by requiring
that the equilibrium state shares the same
expected energy,
particle numbers,
etc.~as the actual initial state of the bath.

For example, if each bath has a
 grand canonical reference state,
Eq.~\eqref{eq:GenEFdef} reduces to
the
familiar form~\cite{deGr84, Kond14, Espo10a, Ptas19a}:
\begin{align}
\EF
= \sum_{b \in \mathbb{B}} \frac{Q^{(b)}}{T^{(b)}} - \frac{1}{T^{(b)}} \sum_{\ell} \mu^{(b, \ell)} \Delta \braket{N^{(b, \ell)}} ~,
\label{eq:SenvDef1}
\end{align}
where the heat $Q^{(b)} = \Delta \tr(\rho_t^{(b)} H^{(b)})$
is the expected energy change of
bath $b$
over the course of the process and
$\Delta \braket{N^{(b, \ell)}} = \Delta \tr(\rho_t^{(b)} N^{(b, \ell)}) $ is the expected change in the bath's number of $\ell$-type particles.
Eq.~\eqref{eq:SenvDef1} has been used to
explore entropy production even in the case of arbitrarily small baths~\cite{Espo10a, Ptas19a}.

\section{An Equality for Entropy Production}

In App.~\ref{sec:InfoTheoreticEP}, we combine
Eqs.~\eqref{eq:NetUnitaryEvolution1}--\eqref{eq:GenEFdef}
to find a new information-theoretic expression for
entropy production:
\begin{align}
\tfrac{1}{\kB} \EP = \Delta \, \mathcal{I}(\rho_t ; \rho_t^\text{env}) + \Delta \text{D} [ \rho_t^\text{env} \| \stationary^\text{env} ] ~,
\label{eq:InfoEPwGenericEnvState}
\end{align}
in terms of the quantum
relative entropy
D$[\rho \| \omega] \equiv \tr(\rho \ln \rho) - \tr(\rho \ln \omega)$
and the quantum
mutual information
$\mathcal{I}(\rho_t; \rho_t^\text{env}) = \text{D}[ \rho_t^\text{tot} \| \rho_t \otimes \rho_t^\text{env} ]$.
This can be rewritten as
\begin{align}
\tfrac{1}{\kB} \EP = \Delta \mathcal{T}_t + \sum_{b \in \mathbb{B}} \Delta \text{D}[\rho_t^{(b)} \| \stationary^{(b)}] ~,
\label{eq:EPasCorrPlusNAFE}
\end{align}
which
tells us that entropy production summarizes both the change in total correlation among system and baths 
$\mathcal{T}_t = \text{D} \bigl[ \rho_t^\text{tot} \big\| \rho_t \otimes \bigl( \bigotimes_{b \in \mathbb{B}} \rho_t^{(b)} \bigr) \bigr]$  
as well as the change
in each bath's nonequilibrium addition to free energy $\text{D}[\rho_t^{(b)} \| \stationary^{(b)}]$.

Eqs.~\eqref{eq:InfoEPwGenericEnvState} and \eqref{eq:EPasCorrPlusNAFE}
quantify the entropy production for any quantum process.
Eq.~\eqref{eq:EPasCorrPlusNAFE} significantly
generalizes Ref.~\cite{Espo10a}'s main result, since it allows for any initial conditions of the system and environment, 
with any possible initial correlations between system and environment. 
By the non-negativity of entropy, relative entropy, and mutual information,
Eqs.~\eqref{eq:InfoEPwGenericEnvState} and \eqref{eq:EPasCorrPlusNAFE}
provide a number of new bounds on entropy production and entropy flow
that generalize both the Second Law of Thermodynamics and Landauer's bound.

Notably, either 
(i) 
initial correlation with (or within) the environment
or 
(ii) 
initially nonequilibrium baths
can be `consumed' to
allow anomalous entropy flow---like heat flow from cold to hot baths---against
Second-Law guided intuition~\cite{Hilt11, Jevt12, Mica19, Mica20}.
These negative-entropy-production events are fully accounted for by 
Eqs.~\eqref{eq:InfoEPwGenericEnvState} and \eqref{eq:EPasCorrPlusNAFE}.
They also serve as
a reminder that the Second Law 
has limited validity, based on stricter assumptions.

Under the common assumptions
1)
that the environment begins in
local equilibrium
(which forces $\text{D} [ \rho_0^\text{env} \| \stationary^\text{env} ] = 0$)~\cite{Mica19, Mica20, Espo10a, Jevt12, Reeb14, Ptas19a, Timp20},
and
2)
that
the system and baths are initially uncorrelated (which forces $\mathcal{I}(\rho_0 ; \rho_0^\text{env}) = 0$)~\cite{Espo10a, Reeb14, Ptas19a, Timp20},
we recover the Second Law of thermodynamics $\EP \geq 0$
and the corresponding
Landauer bound $\EF \geq - \Delta S(\rho_t)$.
Entropy production $\EP$ then quantifies
effectively irreversible dissipation beyond Landauer's bound.
However, even when these assumptions are not valid,
entropy production is still useful via its relation to entropy flow---which can, for example, tell us the heat required for any transformation or computation.

\section{Initial-state dependence}

We can now consider how the initial state of the system
affects entropy production.
Varying the initial state of a system,
while holding its initial environment fixed,
enforces an initial product state:
\begin{align}
\rho_0^\text{tot} &= \rho_0 \otimes \rho_0^{\text{env}} ~.
\label{eq:InitProductState1}
\end{align}
The reduced final state of the system is
$\rho_\tau = \QProcess(\rho_0) = \trenv ( \rho_\tau^\text{tot} )$.
Via Eqs.~\eqref{eq:InitProductState1} and \eqref{eq:NetUnitaryEvolution1}
and Stinespring's dilation theorem,
the quantum channel $\QProcess( \cdot )$
can implement any quantum operation on the system~\cite{Stin55}.

We will show that Eq.~\eqref{eq:MainResult1} is
a consequence of Eqs.~\eqref{eq:NetUnitaryEvolution1}--\eqref{eq:GenEFdef}
and
\eqref{eq:InitProductState1}.
Accordingly, Eq.\ \eqref{eq:MainResult1}
applies
to any process,
for any initial state of the system, for any initial state of the environment, and for arbitrarily small system and baths.

\subsection{Derivation of Main Result}

The initial density matrix $\rho_0$
can be represented in an arbitrary orthonormal basis as
$\rho_0 = \sum_{j, k} c_{j, k} \ket{j} \bra{k} $
with  $c_{j, k} = \bra{j} \rho_0 \ket{k}$.
We can
consider all possible variations of the initial density matrix
via changes in these
$c_{j, k}$ parameters.

We aim to expose the $\rho_0$-dependence of entropy production.
From Eq.~\eqref{eq:GenEFdef},
the only $\rho_0$ dependence in $\EF$
is linear via $\rho_\tau^\text{env}$.
Meanwhile, utilizing the spectral theorem,
it is useful to rewrite the change in system entropy as
$\tfrac{1}{\kB} \Delta S(\rho_t)
= - \Delta \sum_{\lambda_t \in \Lambda_{\rho_t}} \lambda_t \ln \lambda_t$,
where
$\Lambda_{\rho_t}$ is the collection of $\rho_t$'s eigenvalues.
We then calculate
the infinitesimal perturbations
$\frac{\partial \lambda_0}{ \partial c_{j, k} } $
and
$\frac{\partial \lambda_\tau}{ \partial c_{j, k} } $.
This leads to an analytic expression for
the partial derivative
$\frac{\partial}{ \partial c_{j, k} }  \EP$.
To consider the consequences of arbitrary variations in the initial density matrix,
we construct a type of gradient
$\grad \EP \equiv \sum_{j, k} \ket{k} \bra{j} \frac{\partial}{\partial c_{j, k}} \EP$
with a scalar product ``$\cdot$'' that gives a type of directional derivative:
$\gamma \cdot \grad \EP
\equiv
\tr( \gamma \grad \EP )$.

For any two density matrices,
$(\rho_0 - \rho_0') \cdot \grad \EP[\rho_0']$
gives the linear approximation of the change
in entropy production
(from the gradient evaluated at $\rho_0'$)
if we were to change the initial density matrix from $\rho_0'$
towards $\rho_0$.
Notably,
using our directional derivative this way
allows us to stay along the manifold of density matrices (due to the convexity of quantum states),
while inspecting the effect of all possible infinitesimal changes to $\rho_0'$.

{\Lem \label{lem:MismatchDot}
For any two density matrices $\rho_0$ and $\rho_0'$:
\scalebox{0.96}{
$\!\!\!$
\begin{minipage}{1.03\columnwidth}
\begin{align*}
\tfrac{1}{\kB}
\rho_0 \cdot \! \grad \EP[\rho_0']
&=
\tr( \rho_0 \ln \rho_0' ) - \tr( \rho_\tau \ln \rho_\tau' )
- \tr( \rho_\tau^\text{env} \ln \stationary^\text{env} )
~.
\end{align*}
\end{minipage}
}
}

See
App.~\ref{sec:GradientDerivation}
for further details of the derivation.
Hence,
for any initial density matrix:
$\rho_0 \cdot \grad \EP
=
\EP - \kB
\tr( \rho_0^\text{env} \ln \stationary^\text{env} )
$.

It is worthwhile to consider any density matrix $\q_0$ that would
lead to minimal entropy production under the control protocol $\drive$:
\begin{align}
\q_0
\in
\argmin_{\rho_0} \EP ~.
\end{align}
If
(on the one hand)
$\q_0$ has full rank,
it must be true that
\begin{align}
(\rho_0 - \q_0) \cdot \grad \EP[\q_0] = 0
\label{eq:ExtremumProperty}
\end{align}
for any density matrix $\rho_0$.
I.e., moving from $\q_0$ infinitesimally in the direction of any other initial density matrix cannot
produce a linear change in the dissipation.
Expanding Eq.~\eqref{eq:ExtremumProperty},
$\rho_0 \cdot \grad \EP[\q_0] - \q_0 \cdot \grad \EP[\q_0] = 0$,
according to Lemma \ref{lem:MismatchDot}
yields our main result:
{\The \label{thm:MainThm}
If $\q_0 \in
\argmin_{\rho_0} \EP$ has full rank, then
\begin{align}
\EP - \EP[\q_0] = \kB \text{D} \bigl[ \rho_0 \big\| \q_0 \bigr] - \kB \text{D} \bigl[ \rho_\tau \big\| \q_\tau \bigr] ~.
\label{eq:MainResult}
\end{align}
If (on the other hand)
$\argmin_{\rho_0} \EP$ has a nontrivial nullspace,
then Eq.~\eqref{eq:MainResult} can be extended
by supplementing  $\q_0$ with the
successive minimally dissipative density matrices on the nullspace.
This extension of our main result is derived and discussed in
Appendix \ref{sec:Genq0}.
}

Relative entropy quantifies distinguishability between two quantum states
upon the most discerning
hypothesis-testing measurements~\cite{Hiai91, Vedr02}.
As an immediate consequence of Eq.~\eqref{eq:MainResult},
a state $\rho_0$ dissipates minimally ($\EP[\rho_0] = \EP[\q_0]$) if it
retains all distinction from
a minimally dissipative state $\q_0$ with the same support;
i.e., if  $\text{D} \bigl[ \rho_0 \big\| \q_0 \bigr] - \text{D} \bigl[ \Gamma ( \rho_0) \big\| \Gamma (\q_0) \bigr] = 0$.
As a consequence,
only unitary channels $\QProcess(\rho_0) = U \rho_0 U^\dagger$
can achieve minimal dissipation via a single protocol for all initial states (since only unitary and antiunitary transformations preserve relative entropy~\cite{Moln10} and the latter are not physical~\cite{Buvz99}).

{\Cor \label{thm:Irreversibility}
Nonunitary
operations
cannot be thermodynamically optimized for all initial states simultaneously.
}

Recall that any
transformation of a
system that begins uncorrelated with equilibrium baths
satisfies the Second Law ($\EP \geq 0$) and Landauer's bound ($\EF \geq - \Delta S$).
Cor.~\ref{thm:Irreversibility}
then says that
any such \emph{nonunitary} transformation
requires dissipation
\emph{beyond} Landauer's bound
for at least some input states.
To refine this corollary,
App.~\ref{sec:UnitarySubspaces},
clarifies the thermodynamic implication of
locally unitary
subspaces of
nonunitary transformations.

It is interesting to compare Eq.~\eqref{eq:MainResult} to the classical result by Kolchinsky and Wolpert~\cite{Kolc17},
which gives this dissipation in terms of the Kullback--Leibler divergence $\text{D}_\text{KL}$ instead of the quantum relative entropy D.
The classical probability distribution
$\mathcal{P}_t$
induced by projecting $\rho_t$ onto $\q_t$'s eigenbasis
(at any time $t$)
can be represented as
\begin{align}
P_t \equiv
\sum_{s \in \Lambda_{\q_t}}
\mathcal{P}_t(s)
\ket{s} \bra{s}
~,
\label{eq:ClassicalProjection}
\end{align}
where each $\ket{s}$ is an eigenstate of $\q_t$, and $\mathcal{P}_t(s) = \braket{s | \rho_t | s}$.
Since $P_t$ and $\q_t$ are diagonal in the same basis,
the relative entropy $\text{D} \bigl[ P_t \big\| \q_t \bigr]$
reduces to the Kullback--Leibler divergence $\text{D}_\text{KL} \bigl[ \mathcal{P}_t \big\| \mathcal{Q}_t \bigr]$,
where
$\mathcal{Q}_t(s) = \braket{s | \q_t | s}$.
However the \emph{actual} density matrix
$\rho_t$ is typically not diagonalized by $\q_t$'s eigenbasis---but
rather exhibits coherence there.
This coherence is naturally quantified by the so-called `relative entropy of coherence'~\cite{Baum14}
\begin{align}
C_{\q_t}(\rho_t)
&=
\tr( \rho_t \ln \rho_t ) - \tr ( P_t \ln P_t )
~.
\end{align}

As shown in Appendix \ref{sec:RelEntDecomp},
the extra dissipation from starting with the density matrix $\rho_0$ rather than the minimally dissipative $\q_0$
is given for any finite-duration nonequilibrium transformation
as
\begin{align}
\tfrac{1}{\kB} ( \EP - \EP[\q_0] )
&=
 \text{D}_\text{KL} \bigl[ \mathcal{P}_0 \big\| \mathcal{Q}_0 \bigr] -  \text{D}_\text{KL} \bigl[ \mathcal{P}_\tau \big\| \mathcal{Q}_\tau \bigr]
 \nonumber \\
& \qquad \! + \,
 C_{\q_0}(\rho_0) - C_{\q_\tau}(\rho_\tau)
 ~.
\label{eq:EPasWKplusCoherence}
\end{align}
We see that the quantum correction to the classical dissipation
is exactly the
change of
coherence on the minimally dissipative eigenbasis.

\section{Several important cases}
\label{sec:Important_cases}

We now consider several important cases.

\subsection{Relaxation to equilibrium}

We
first consider the case
of time-independent driving and
weak coupling to a single bath that can exchange
energy and possibly particles, volume, etc.
There is zero dissipation if the system starts in the equilibrium state $\stationary_x$, so $\q_0 = \stationary_x$
with $\EP[\stationary_x ] = 0$.
The dissipation when starting in state $\rho_0$ is thus
\begin{align}
\EP
=
\kB \text{D} \bigl[ \rho_0 \big\| \stationary_x \bigr] -  \kB \text{D} \bigl[ \rho_\tau \big\| \stationary_x \bigr]
~.
\label{eq:Relax}
\end{align}
In the case of a canonical thermal bath at temperature $T$,
this reduces to a well known result
since $\mathcal{F}_t^\text{add} = \kB T \text{D} \bigl[ \rho_t \big\| \stationary_{x_t} \bigr] $ is then the
system's nonequilibrium addition to free energy~\cite{Deff11, Bran13, Aabe14, Nara19, Scan19}.
Notably, Eq.~\eqref{eq:Relax} applies more generally to
equilibration in \emph{any} thermodynamic potential.
When the equilibrium state $\stationary_x$ is diagonal in the energy eigenbasis,
this dissipation can be
attributed to
the change
in probability of the system's energy eigenstates
$\text{D}_\text{KL} \bigl[ \mathcal{P}_0 \big\| \stationary_x \bigr] -  \text{D}_\text{KL} \bigl[ \mathcal{P}_\tau \big\| \stationary_x \bigr] $
and the decoherence in the energy eigenbasis
$C_{\stationary_x}(\rho_0) - C_{\stationary_x}(\rho_\tau)$.

\begin{figure}[t]
\begin{center}
\includegraphics[width=1.0\columnwidth]{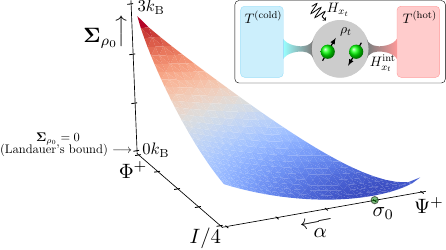}
\end{center}
\caption{Whether preparing a Bell state, erasing quantum memory, or extracting work,
no control protocol can be simultaneously thermodynamically optimal for all initial quantum states on which it operates.
Inset:
A quantum system $\rho_t$ of two qubits is driven by a time-dependent Hamiltonian $H_{x_t}$ and time-dependent interaction $H_{x_t}^\text{int}$ with thermal reservoirs.  Main:
Suppose the control protocol $\drive$
resets the two qubits in finite time,
and achieves minimal dissipation
when operating on the
\emph{noisy Bell}
state:
$\q_0 = \tfrac{1-\alpha}{2} \bigl( \ket{01} + \ket{10} \bigr)
\bigl( \bra{01} + \bra{10} \bigr) + \tfrac{\alpha}{4} I $,
with some particular $\alpha \in [0, 1]$.
If the reset protocol is designed to be optimal for erasing
the $\Psi^+$ Bell state ($\alpha=0$),
then the same protocol approaches \emph{infinite dissipation} as $\rho_0$ approaches any of the
$\Phi^+$, $\Phi^-$, or $\Psi^-$
Bell states
orthogonal to it.
For $0 < \alpha \ll 1$, the dissipation  $\EP$ scales as $\ln(1/\alpha)$ near these three Bell states.
If $\alpha=1$---i.e., the reset protocol is optimized for erasing
completely randomized classical bits---then \emph{any} initial quantum state can be
erased with no more than $\kB \ln 2$ of dissipation beyond $\EP[\q_0]$
(where $\EP[\q_0]$ can be engineered to be arbitrarily small).
This dissipation is distinct and additive to the Landauer cost of erasure.
Dissipation is plotted over a triangular domain, 
which is only a subset of possible input states.
This triangular domain is the convex hull of input states $I/4$, $\Phi^+$, and $\Psi^+$.
	}
\label{fig:Setup}
\end{figure}

\subsection{Reset}

Consider
any control protocol $\drive$ that implements
\texttt{RESET}
to a desired state
$r_\tau$
from all
initial quantum states $\rho_0$.~\footnote{This indeed requires
an open quantum system, which introduces the possibility of dissipation.}
For example: to erase
any number of
qubits (or qutrits, etc.) or, similarly,
to initialize an entangled Bell state.
Reliable \texttt{RESET} protocols will have
a small error tolerance $\epsilon$
that upper bounds the
trace distance between the desired state $r_\tau$ and
the actual final state $\rho_\tau$,
for any initial state $\rho_0$.

It should be noted that 
the desired state $r_\tau$ may
differ from the minimally dissipative state $\q_0$.  Meanwhile, the minimal
dissipation $\EP[\q_0]$ may be non-zero.

Appendix \ref{sec:ResetBounds}
shows that
$\text{D} \bigl[ \rho_\tau \big\| \q_\tau \bigr]
\leq  4\epsilon \ln \Bigl( \frac{d}{4 \epsilon \sqrt{s_\text{min}}} \Bigr) $,
where $d$ is the dimension of the Hilbert space and
$s_\text{min}$ is the smallest eigenvalue of $\q_\tau$.
In the $\epsilon \to 0$ limit of high-fidelity \texttt{RESET},
$\text{D} \bigl[ \rho_\tau \big\| \q_\tau \bigr] \to 0$, and
we obtain
\begin{align}
\tfrac{1}{\kB} ( \EP - \EP[\q_0] )
=  \text{D} \bigl[ \rho_0 \big\| \q_0 \bigr] ~.
\label{eq:ResetDiss}
\end{align}
When the Second Law is valid,
this
implies that
implementing erasure with the fixed protocol $\drive$
must result in entropy production for any
$\rho_0 \neq \q_0$.

For thermodynamic efficiency, the reset protocol
must be designed around the expected initial state.
But what if the initial state is unknown, or expectations are misaligned?
Fig.~\ref{fig:Setup} illustrates the thermodynamic cost of misaligned expectations
when the \texttt{RESET} operation is applied to two qubits
that the protocol is not optimized for.
This exposes
the risk of divergent dissipation upon overspecialization---when the protocol operates on a state that is nearly orthogonal to the anticipated initial state $\q_0$.

For energetic efficiency in multiple use cases---say in resetting unknown
qubits of a quantum computer---it
is advisable when constructing such protocols to hedge thermodynamic bets.
Bringing $\q_0$ closer to the fully mixed state (proportional to the identity) assures that
no state is orthogonal to it.

\subsection{Thermodynamic Cost of Modularity}

Just as we break a classical circuit into elementary logic gates, building complex quantum operations generally involves decomposing them into 
elementary operations on smaller subsystems. The advantage of this modular design is that such elementary operations can then be mass produced,
and then arranged to perform any number of desired complex tasks. Our main result implies that this modularity also incurs a thermodynamic
price. Notably, each elementary operation must be optimized individually, without prior knowledge of where they'll be placed in the grander scale. 
Accordingly, a mismatch is expected 
between the optimal
vs.\ actual inputs, and
dissipation
is often unavoidable.

Consider a collection of elementary quantum operations $G_n$, acting on respective Hilbert space $\mathcal{H}_n$. Each is at-best individually optimized,
such that dissipation is minimized for some $\q_{0,n}$. Suppose we place them in parallel to build a composite $N$-partite operation $\bigotimes_{n=1}^N G_n$.
Individual optimization implies that the minimally dissipative state, $\q_0 = \bigotimes_{n=1}^N \q_{0, n}$
will take a product form. 
Each reduced state $\rho_{0,n}$
input to the elementary operation $G_n$
may produce some mismatch dissipation
if $\rho_{0,n} \neq \q_{0,n}$.
Moreover,
if the input to our computation, $\rho_0$, has correlations, we will incur additional dissipation from any lost correlation: 
\begin{align}
\tfrac{1}{\kB} \bigl( \EP - \EP[\otimes_n \rho_{0,n}] \bigr)
=
- \Delta \text{D} \bigl[ \rho_t \big\| \bigotimes_{n=1}^N \rho_{t, n } \bigr] ~.
\label{eq:ModularityDiss1}
\end{align}
(See App.~\ref{sec:ModDiss} for details of the derivation.)
When $N=2$, this corresponds to the change in quantum mutual information induced by $G_1 \otimes G_2$ 
~\cite{Boyd18b, Riec19b, Loom20a}.
This quantity is relevant, for example, when correlations are destroyed through local measurements or local erasure protocols. 

In contrast to previous approaches~\cite{Boyd18b, Riec19b, Loom20a}, our derivation of modularity dissipation 
is valid for any parallel computation occurring in finite time, 
regardless of the local free energies of the memory elements utilized.

Individual component optimization 
is further limited
when
these parallel operations 
are embedded in series---since 
gate placement
in a larger circuit will
almost invariably change the distribution of input states it will encounter. 
Thus, a complex composite circuit is very unlikely to reach Landauer's limit.

\section{Generalization to other optimization problems}
\label{sec:OtherOptProblems}

Our main result appears superficially similar to a recent result by Kolchinsky et al.~\cite{Kolc17b},
which describes the initial-state dependence of nonequilibrium free energy gain.
In Ref.~\cite{Kolc17b},
it was shown that the maximal
nonequilibrium free energy gain $\Delta \mathcal{F}_{\chi_t}$
differs from the free energy gain from any initial state $\rho_0$ according to the contraction of relative entropy between the two states $\rho_t$ and $\chi_t$ over the course of the protocol:
\begin{align}
\Delta \mathcal{F}_{\chi_t} - \Delta \mathcal{F}_{\rho_t} = \kB T \text{D} [ \rho_0 \| \chi_0] - \kB T \text{D} [ \rho_\tau \| \chi_\tau]~.
\label{eq:MaxFreeEnergyContraction}
\end{align}
The main results there and here are nevertheless distinct since
the initial state that leads to maximal free energy gain is
typically \emph{not} the same as the initial state that leads to minimal dissipation.
Yet the similarity of the two results suggests a more general overarching result.
Indeed, we have found a general theorem that contains these results as important cases.

To investigate, how any quantity depends on the initial state of the system, 
presupposes
an initial
product state of the system and environment:
$\rho_0^\text{tot} = \rho_0 \otimes \rho_0^\text{env}$.  
Furthermore, suppose that the joint system and environment evolve according to some unitary time evolution, 
such that the reduced state at time $\tau$ is given by
$\rho_\tau = \tr_{\text{env}} \bigl( \mathcal{U} \rho_0 \otimes \rho_0^\text{env} \mathcal{U}^\dagger \bigr)$.

Consider any real-valued functional of the initial density matrix:
\begin{align}
f(\rho_0) = a(\rho_0) + \tr \bigl( \rho_0 \ln \rho_0 - \rho_\tau \ln \rho_\tau \bigr) ~,
\end{align}
and its minimizer:
\begin{align}
\alpha_0 \in \argmin_{\rho_0} f(\rho_0) ~.
\end{align}
Suppose that
$a(\rho)$ is an affine function, such that it can be written as
$a(\rho) = \ell(\rho) + c$, where $\ell(\rho)$ is a linear function of $\rho$ and $c$ is a constant.

{\The \label{thm:GenLinYieldsRelEnt}
If $a(\rho)$ is
an affine
function of $\rho$, 
and
$\alpha_0 \in \argmin_{\rho_0} f(\rho_0)$ has a trivial nullspace,
then
$f(\rho_0) - f(\alpha_0) =
\text{D} \bigl[ \rho_0 \big\| \alpha_0 \bigr] - \text{D} \bigl[ \rho_\tau \big\| \alpha_\tau \bigr] $.
}

Our proof of this theorem, given in App.~\ref{sec:RelatedOptimization}, 
parallels the proof of Thm.~\ref{thm:MainThm}.
If $\alpha_0$ has a non-trivial nullspace, then Thm.~\ref{thm:GenLinYieldsRelEnt} can be extended
as done in App.~E.

The classical limit of Thm.~\ref{thm:GenLinYieldsRelEnt} is closely related
to a result recently derived in Ref.~\cite[Thm.~1]{Wolp19}.

Our main result concerns entropy production:
$f(\rho_0) = \tfrac{1}{\kB}  \EP$,
for which
$a(\rho_0) = \tr( \rho_0^\text{env} \ln  \stationary^\text{env} )
- \tr \bigl( \trsys ( \mathcal{U}_{\drive}  \rho_0 \otimes \rho_0^\text{env}  \mathcal{U}_{\drive}^\dagger ) \ln  \stationary^\text{env} \bigr)$
is indeed an affine function with $c = \tr( \rho_0^\text{env} \ln  \stationary^\text{env} )$.

Ref.~\cite{Kolc17b} considers the (negative) change in nonequilibrium free energy $f(\rho_0) = \beta \mathcal{F}_{\rho_0} - \beta \mathcal{F}_{\rho_\tau} $, for which $a(\rho_0) = \beta \tr(\rho_0 H_0) - \beta \tr(\rho_\tau H_\tau) $ is a linear function of $\rho_0$.
(Note that the minimum of the decrease in nonequilibrium free energy is the maximum increase.)
With this form of $a(\rho)$, Thm.~\eqref{thm:GenLinYieldsRelEnt}
immediately yields Eq.~\eqref{eq:MaxFreeEnergyContraction}.

Another possibility is obtained if we simply let $a(\rho) = 0$.
Then we find a new result about the initial-state dependence of entropy change:
\begin{align}
\Delta S(\rho_t) - \Delta S(\kappa_t)
= \kB \text{D} \bigl[ \rho_0 \big\| \kappa_0 \bigr] - \kB \text{D} \bigl[ \rho_\tau \big\| \kappa_\tau \bigr] ~,
\end{align}
where $\kappa_0 \in \argmin_{\rho_0} S(\rho_\tau) - S(\rho_0) $.

A simple example reveals that these optimization problems indeed have different solutions (i.e., different $\alpha_0$s).
Consider a double-well energy landscape.  The right well is raised in a very short duration $\tau$
in which the system cannot fully relax.
The initial distribution that minimizes dissipation primarily occupies the left well.
The initial distribution that maximizes free energy gain primarily occupies the right well.

\section{Discussion}

In Eqs.~\eqref{eq:EPasCorrPlusNAFE} and \eqref{eq:MainResult},
we exactly 
quantify dissipation in
finite-time transformations of open quantum systems, and identify new relationships among dissipation, 
correlation,
and distinction.  When 
a system begins
in any state
other than the
minimally dissipative
initial state,
the extra dissipation is exactly
the contraction of the quantum relative entropy
between them over the duration of the control protocol---their loss of distinguishability.

This has immediate consequences for thermally efficient quantum information processing.
Crucially,
a quantum control protocol cannot generally be made
thermodynamically optimal for all possible input states,
creating unavoidable dissipation beyond Landauer's in quantum state preparation.
Meanwhile, it imposes extra thermodynamic cost to modular computing architectures,
where one wishes to optimize the thermal efficiency of certain quantum operations
without pre-knowledge of how they will fit within a composite quantum protocol.

Our results are relevant for both quantum and classical finite-time thermodynamics.
In the quantum regime, we have shown that the loss of coherence on the minimally dissipative eigenbasis
directly contributes to dissipation.
Appendix \ref{sec:NonselectiveMeas} further considers decoherence, and
shows how our results quantify dissipation associated with non-selective measurement.
Other applications of our finite-time equalities for nonequilibrium thermodynamics
remain to be explored.
For example,
App.~\ref{sec:NESS} suggests
that our results may be leveraged
to analyze
dissipation in
relaxation to
nonequilibrium steady states.

Our general framework has allowed the investigation of entropy production's dependence on initial conditions,
even beyond the domain of the Second Law's applicability.
The role of system--environment correlation and nonequilibrium baths was highlighted in Eq.~\eqref{eq:EPasCorrPlusNAFE}. 
The initial properties of the system itself were then shown to have many important consequences,
through the implications of Eq.~\eqref{eq:MainResult}.
From a broad philosophical perspective,
these lessons extend
our understanding of effective irreversibility in quantum mechanics, despite its global unitarity.
More specifically, we 
have found how
energetic resources are taxed for
coherence,
correlations, and
misaligned expectations.

\acknowledgements

We are grateful to
Felix Binder,
Alec Boyd,
Artemy Kolchinsky,
Gabriel Landi,
Varun Narasimhachar,
and David Wolpert
for useful discussions relevant to this work.
This work was supported by the
National Research Foundation and
L'Agence Nationale de la Recherche joint Project No.~NRF2017-NRFANR004 VanQuTe,
the National Research Foundation of Singapore Fellowship No.~NRF-NRFF2016-02,
the Singapore Ministry of Education Tier 1 grant RG190/17,
and the FQXi Grant `Are quantum agents more energetically efficient at making predictions?'.

\appendix

\section{Information-theoretic equalities for entropy production when the system begins correlated with nonequilibrium environments}
\label{sec:InfoTheoreticEP}

Here, we derive new information-theoretic
equalities 
(Eqs.~\eqref{eq:InfoEPwGenericEnvState} and \eqref{eq:EPasCorrPlusNAFE})
for entropy production
in the general case that allows for arbitrary initial correlation between the system and nonequilibrium environments.
This generalizes a number of related results~\cite{Espo10a, Jevt12, Reeb14, Ptas19a, Mica19, Mica20}.

Our result follows from
the definition of entropy production together with the
assumption of partially controlled unitary time evolution of the joint system and environment.
Recall that entropy production is defined as
\begin{align}
\EP \equiv \EF + \Delta S(\rho_t) ~,
\label{eq:EPdef_forApp}
\end{align}
with
\begin{align}
\Delta S(\rho_t) = \kB \tr(\rho_0 \ln \rho_0) - \kB \tr(\rho_\tau \ln \rho_\tau)
\end{align}
and
\begin{align}
\EF \equiv
\kB  \tr(\rho_0^\text{env} \ln \stationary^\text{env})
- \kB  \tr(\rho_\tau^\text{env} \ln \stationary^\text{env})  ~.
\label{eq:GenEFdef_forApp}
\end{align}
Notably, the assumption of joint unitary dynamics
\begin{align}
\rho_{\tau}^\text{tot} = \mathcal{U}_{\drive} \rho_0^\text{tot}
\mathcal{U}_{\drive}^\dagger ~,
\end{align}
guarantees that the net von Neumann entropy of the system and baths remains unchanged by the protocol:
\begin{align}
\Delta S(\rho_t^\text{tot}) = 0 ~.
\label{eq:ZeroNetEntropyChange_forApp}
\end{align}

Recall that in general, the joint entropy between system and environment can be decomposed as
\begin{align}
S(\rho_t^\text{tot}) = S(\rho_t) + S(\rho_t^\text{env}) - \kB \mathcal{I}( \rho_t ; \rho_t^\text{env})   ~.
\end{align}
Combining this decomposition with Eq.~\eqref{eq:ZeroNetEntropyChange_forApp},
we find
\begin{align}
\Delta S(\rho_t^\text{tot}) = \Delta S(\rho_t) + \Delta S(\rho_t^\text{env}) - \kB \Delta \mathcal{I}( \rho_t ; \rho_t^\text{env})  = 0 ~,
\label{eq:vNEntropyDecomp}
\end{align}
which can be rearranged as
\begin{align}
\Delta S(\rho_t) = - \Delta S(\rho_t^\text{env}) + \kB \Delta \mathcal{I}( \rho_t ; \rho_t^\text{env})   ~.
\label{eq:RewritingDeltaS}
\end{align}
Combining Eqs.~\eqref{eq:EPdef_forApp} and \eqref{eq:RewritingDeltaS},
we find that
\begin{align}
\tfrac{1}{\kB} \EP =  \kB \Delta \mathcal{I}( \rho_t ; \rho_t^\text{env}) + \EF - \Delta S(\rho_t^\text{env}) ~.
\label{eq:EP_intermediaryresult}
\end{align}

Now combining Eqs.~\eqref{eq:EP_intermediaryresult} and \eqref{eq:GenEFdef_forApp},
we obtain
\begin{align}
\EP =  \Delta \mathcal{I}( \rho_t ; \rho_t^\text{env}) + \Delta \text{D} [ \rho_t^\text{env} \| \stationary^\text{env} ] ~.
\label{eq:EP_InfoTheoryDecomp}
\end{align}

By the non-negativity of entropy, relative entropy, and mutual information,
Eq.~\eqref{eq:EP_InfoTheoryDecomp}
provides a number of new bounds on entropy production and entropy flow
that generalize both the Second Law of Thermodynamics and Landauer's bound.

Initial correlation between system and environment can be thought of informally as catching a Maxwellian demon mid act, after correlation has already been established.  To bring about correlation requires resources but, if it is already present, correlation can be consumed to reduce entropy production.
Similarly, initially nonequilibrium environments can act as thermodynamic resources to reduce entropy production.

Alternatively, Maxwell's demons can be accounted for within the standard framework of the Second Law by including the demon and `system under study' as two subsystems of a larger system embedded in an initially equilibrium environment (which is, furthermore, initially uncorrelated with the two subsystems)~\cite{Boyd16, Boyd15a, Riec19b}.  But our generalized equality for entropy production allows for the boundaries between system and environment to be drawn arbitrarily, and thus describes entropy production (as a function of those chosen boundaries) in a broader set of scenarios.

The consequences of initial correlation or initially nonequilibrium environment can be quite astonishing.  For example, either feature allows for heat to reliably flow from hot to cold reservoirs, as in Ref.~\cite{Mica19}.

Because of the product structure of the local-equilibrium reference state,
Eq.~\eqref{eq:EP_InfoTheoryDecomp} can be rewritten as
\begin{align}
\EP
&= \Delta S(\rho_t)
- \Delta \tr(  \rho_t^\text{env} \ln \stationary^\text{env} )
\\
&= \Delta S(\rho_t)
+ \sum_{b \in \mathbb{B}} - \Delta \tr( \rho_t^{(b)} \ln \stationary^{(b)} )  \\
&=
\Bigl[ \Delta S(\rho_t) + \sum_{b \in \mathbb{B}} \Delta S(\rho_t^{(b)}) \Bigr]
+ \sum_{b \in \mathbb{B}} \Delta \text{D}[ \rho_t^{(b)} \| \stationary^{(b)} ]  \\
&=
\Delta \mathcal{T}_t  + \sum_{b \in \mathbb{B}} \Delta \text{D} [ \rho_t^{(b)} \| \stationary^{(b)} ]   ~,
\label{eq:TotCorrDecompOfEP}
\end{align}
where $\mathcal{T}_t = \text{D} \bigl[ \rho_t^\text{tot} \big\| \rho_t \otimes \bigl( \bigotimes_{b \in \mathbb{B}} \rho_t^{(b)} \bigr) \bigr]$ is the \emph{total correlation} among system and baths.
(Here and throughout, $\rho_t^{(b)} = \trallbutb(\rho_t^\text{tot})$ is the reduced state of bath $b$ at time $t$.)
This allows the perspective that
entropy production is the change in the nonequilibrium addition to free energy of each bath, plus the change in total correlation among the system and all baths.
Ref.~\cite[Eq.~(17)]{Espo10a}
attained a special case of this result,
under the assumption that the baths are initially in equilibrium and uncorrelated from the system and from each other.
Our generalization allows rigorous consideration of
much broader phenomena, in the presence of initial correlation with and between any nonequilibrium baths.

\begin{widetext}
	
\section{Decompositions and interpretations of entropy flow}
\label{App:EF}

As will be made clear in the following, the expected value of
entropy flow can generically be written as $\EF = -\Delta \tr(\rho_t^\text{env} \ln \stationary^\text{env})$.
By adding and subtracting the change in von Neumann entropy of the environment, this can be rewritten as
\begin{align}
\EF
& =
\Delta S(\rho_t^\text{env}) + \kB \Delta \text{D} [\rho_t^\text{env} \| \stationary^\text{env}] ~.
\end{align}

Recall that
$\stationary^\text{env} = \bigotimes_{b \in \mathbb{B}} \stationary^{(b)}$,
where $\stationary^{(b)}$ is the reference equilibrium state of bath $b$.
Using the mathematical fact that $\tr[\rho' \ln (\rho_A \otimes \rho_B)] = \tr[\tr_{\mathcal{H}_B}(\rho') \ln(\rho_A)] + \tr[\tr_{\mathcal{H}_A}(\rho') \ln(\rho_B)]$,
we can alternatively decompose the entropy flow into the contributions from each bath.
We will denote the reduced state of bath $b$ at time $t$ as: $\rho_t^{(b)} = \trallbutb(\rho_t^\text{tot})$.
It then follows that
\begin{align}
\EF
& =
\kB
\tr(\rho_0^\text{env} \ln \stationary^\text{env} )
-  \kB \tr(\rho_\tau^\text{env} \ln \stationary^\text{env} )
\\
&=
\kB
\sum_b
\bigl(
\tr(\rho_0^{(b)} \ln \stationary^{(b)} )
-  \tr(\rho_\tau^{(b)} \ln \stationary^{(b)} )
\bigr)
\label{eq:EF1b}
\\
&=   \sum_b \Delta S(\rho_t^{(b)}) + \kB \Delta \text{D} [ \rho_\tau^{(b)} \| \stationary^{(b)}] ~,
\label{eq:EF2b}
\end{align}
I.e., Entropy flow is the change in von Neumann entropy of the baths, plus the additional nonequilibrium free energy
$\kB \Delta \text{D} [ \rho_t^{(b)} \| \stationary^{(b)}]$
gained by each bath.
Eq.~\eqref{eq:EF2b}
generalizes
other decompositions of the entropy flow
presented in Refs.~\cite{Reeb14, Ptas19a}, which are obtained upon assumption of initially equilibrium baths that are uncorrelated with each other and with the system.
Either Eq.~\eqref{eq:EF1b} or Eq.~\eqref{eq:EF2b} can be taken as the general definition of entropy flow,
which allows our results to apply to baths from any thermodynamic ensemble~\cite{Albe01}.

Entropy flow has been of primary interest
since the origins of thermodynamics.
In its simplest form, when there is a single thermodynamic bath that a system can exchange energy with,
entropy flow takes the form of
$\frac{\text{heat flow to bath}}{\text{temperature of bath}}$.
In quasistatic reversible transformations of an equilibrium system, this entropy flow will be equal to the change in equilibrium entropy of the system.  However, for  general irreversible transformations, the Second Law tells us that the entropy flow exceeds the change in system entropy.
(Our results, however, caution us that this interpretation of the Second Law can be broken when initial correlations with the environment or the initial nonequilibrium nature of the environment are leveraged.)

More generally, we may consider entropy flow among many thermodynamic baths of any thermodynamic potentials.  For example, baths of different temperatures may be introduced.  Furthermore, chemical potentials and other coarse features of each bath may be incorporated into its thermodynamic potential~\cite{Albe01}.

As an important example, suppose
each bath has a
grand canonical reference state,
such that
\begin{align}
\stationary^{(b)} =
e^{-\tfrac{1}{\kB T^{(b)}} \bigl( H^{(b)} - \sum_\ell \mu^{(b, \ell)} N^{(b, \ell)} \bigr) } / \mathcal{Z}^{(b)} ~,
\label{eq:CanonicalBath}
\end{align}
where
$T^{(b)}$ is the
initial
temperature of the $b^\text{th}$ bath, $H^{(b)}$ is its
Hamiltonian,
$\mu_\ell^{(b)}$
and $N_\ell^{(b)}$ are the
initial
chemical potential and the number operator for the bath's $\ell^\text{th}$ particle type, and
$\mathcal{Z}^{(b)}$ is the grand canonical partition function for the bath~\cite{Reic09}.
Even if the bath is not initially in equilibrium, the initial temperature and initial chemical potentials
are well defined in statistical mechanics
via average energy and expected particle number respectively.
In particular, coarse ``macroscopic'' knowledge about total energy, particle numbers, etc.~gives exactly the correct number of equations to solve for temperature, chemical potentials, etc.  In the grand canonical case,
initial temperature and chemical potentials are fixed by requiring
that the equilibrium state shares the same
expected energy,
$\tr(\stationary^{(b)} H^{(b)}) = \tr(\rho_0^{(b)} H^{(b)})$, and
particle numbers,
$\tr(\stationary^{(b)} N^{(b, \ell)}) = \tr(\rho_0^{(b)} N^{(b, \ell)})$,
as the actual initial state of the bath.

We can now use Eq.~\eqref{eq:EF1b} to rewrite the entropy flow in the case of grand canonical reference states for the bath.
First, we note that
\begin{align}
-\kB \ln \stationary^{(b)} = - \kB \ln(\mathcal{Z}^{(b)}) + \frac{H^{(b)}}{T^{(b)}} - \frac{1}{T^{(b)}} \sum_\ell \mu^{(b, \ell)} N^{(b, \ell)} ~.
\end{align}
Accordingly,
the expected change in this observable is
\begin{align}
-\kB \Delta \tr( \rho_t^{(b)} \ln \stationary^{(b)} ) =   \frac{ \Delta \tr(\rho_t^{(b)} H^{(b)})}{T^{(b)}} - \frac{1}{T^{(b)}} \sum_\ell \mu^{(b, \ell)} \Delta \tr( \rho_t^{(b)} N^{(b, \ell)} ) ~.
\end{align}
Eq.~\eqref{eq:EF1b}
then reduces to
the
familiar form~\cite{deGr84, Kond14, Espo10a, Ptas19a}:
\begin{align}
\EF
= \sum_{b \in \mathbb{B}} \frac{Q^{(b)}}{T^{(b)}} - \frac{1}{T^{(b)}} \sum_{\ell} \mu^{(b, \ell)} \Delta \braket{N^{(b, \ell)}} ~,
\label{eq:SenvDef1b}
\end{align}
where the heat $Q^{(b)} = \Delta \tr(\rho_t^{(b)} H^{(b)})$
is the expected energy change of
bath $b$
over the course of the process and
$\Delta \braket{N^{(b, \ell)}} = \Delta \tr(\rho_t^{(b)} N^{(b, \ell)}) $ is the expected change in the bath's number of $\ell$-type particles.
Eq.~\eqref{eq:SenvDef1b} has been used to
explore entropy flow and entropy production even in the case of arbitrarily small baths~\cite{Espo10a, Ptas19a}.

\section{Derivation of the Gradient}
\label{sec:GradientDerivation}

To vary the initial state of the system while holding the initial state of the environment fixed requires
$\rho_0^\text{tot} = \rho_0 \otimes \rho_0^\text{env} $.
We can fully specify the initial state of the system as $\rho_0 = \sum_{j,k} c_{j,k} \ket{j} \bra{k}$.
From Eqs.~\eqref{eq:NetUnitaryEvolution1}
through \eqref{eq:GenEFdef}
of the main text,
we can then express
the entropy production as
\begin{align}
\tfrac{1}{\kB}  \EP
&=
\tr(\rho_0 \ln \rho_0) - \tr(\rho_\tau \ln \rho_\tau)
+   \tr( \rho_0^\text{env} \ln  \stationary^\text{env} )
- \tr \bigl( \trsys ( \mathcal{U}_{\drive}  \rho_0 \otimes \rho_0^\text{env}  \mathcal{U}_{\drive}^\dagger ) \ln  \stationary^\text{env} \bigr)
\\
&=
\Bigl( \! \sum_{\lambda_0 \in \Lambda_{\rho_0}} \! \! \lambda_0 \ln \lambda_0 \Bigr)
- \Bigl( \! \sum_{\lambda_\tau \in \Lambda_{\rho_\tau}} \!\!  \lambda_\tau \ln \lambda_\tau \Bigr)
+   \tr( \rho_0^\text{env} \ln  \stationary^\text{env} )
-
\sum_{j, k} c_{j, k}
\tr \bigl( \trsys ( \mathcal{U}_{\drive}   \ket{j} \bra{k} \otimes \rho_0^\text{env}  \mathcal{U}_{\drive}^\dagger ) \ln  \stationary^\text{env} \bigr)
 ~.
\label{eq:EP_suitable_for_derivation1}
\end{align}
Varying a single parameter $c_{j, k}$ of the initial density matrix
yields the partial derivative:
\begin{align}
\tfrac{1}{\kB} \frac{\partial}{ \partial c_{j, k} }  \EP
&=
\Bigl( \!\!\! \sum_{\lambda_0 \in \Lambda_{\rho_0}} \! \! \Bigl( \frac{\partial \lambda_0}{ \partial c_{j, k} } \Bigr) \ln \lambda_0 + \frac{\partial \lambda_0}{ \partial c_{j, k} }  \Bigr)
-
\Bigl( \!\!\! \sum_{\lambda_\tau \in \Lambda_{\rho_\tau}} \! \! \Bigl( \frac{\partial \lambda_\tau}{ \partial c_{j, k} } \Bigr) \ln \lambda_\tau + \frac{\partial \lambda_\tau}{ \partial c_{j, k} }  \Bigr)
-
\tr \bigl( \trsys ( \mathcal{U}_{\drive} \! \ket{j} \! \bra{k} \otimes \rho_0^\text{env}  \mathcal{U}_{\drive}^\dagger ) \ln  \stationary^\text{env} \bigr)
 \nonumber
\end{align}
which leads us to evaluate the infinitesimal perturbations
$\frac{\partial \lambda_0}{ \partial c_{j, k} }$ and $\frac{\partial \lambda_\tau}{ \partial c_{j, k} }$
to the eigenvalues of $\rho_0$ and $\rho_\tau$ respectively.

(We could
alternatively choose to vary real-valued variables $c_{j, k}^{(\text{r})}$ and
$c_{j, k}^{(i)}$, such that
$c_{j, k}^{(\text{r})} = \tfrac{1}{2} ( c_{j, k} + c_{k, j} ) $ and
$c_{j, k}^{(i)} = \tfrac{-i}{2} ( c_{j, k} - c_{k, j})$.
Then we can write $\rho_0 = \sum_{j, k } (c_{j, k}^{(\text{r})} + i c_{j, k}^{(i)} ) \ket{j} \bra{k} $
and differentiate with respect to these real variables.
However, it conveniently turns out that $\EP$ is complex-differentiable in all complex-valued $c_{j, k}$ variables, so we can differentiate directly with respect to $c_{j, k}$.)

Starting with the eigen-relation: $\rho_0 \ket{\lambda_0} = \lambda_0 \ket{\lambda_0}$,
we can take the partial derivative of each side:
\begin{align}
\frac{\partial }{ \partial c_{j, k} }  \rho_0 \ket{\lambda_0}
&= \Bigl( \frac{\partial }{ \partial c_{j, k} }  \rho_0 \Bigr) \ket{\lambda_0}
+ \rho_0 \frac{\partial }{ \partial c_{j, k} }  \ket{\lambda_0}
\qquad
=
\qquad
\frac{\partial }{ \partial c_{j, k} }  \lambda_0 \ket{\lambda_0}
= \Bigl( \frac{\partial \lambda_0 }{ \partial c_{j, k} }  \Bigr) \ket{\lambda_0}
+ \lambda_0 \frac{\partial }{ \partial c_{j, k} }  \ket{\lambda_0}
~.
\end{align}
Left-multiplying by $\bra{\lambda_0}$,
and recalling that $\rho_0 = \sum_{j, k } c_{j, k} \ket{j} \bra{k}$,
we obtain
\begin{align}
\Bigl( \frac{\partial \lambda_0 }{ \partial c_{j, k} }  \Bigr) \braket{\lambda_0 |\lambda_0}
=
\frac{\partial \lambda_0 }{ \partial c_{j, k} }
=
\bra{\lambda_0} \Bigl(  \underbrace{ \frac{\partial }{ \partial c_{j, k} }  \rho_0  }_{\ket{j} \bra{k}} \Bigr) \ket{\lambda_0}
+
\underbrace{
	\underbrace{
		\bra{\lambda_0} \rho_0 \frac{\partial }{ \partial c_{j, k} }  \ket{\lambda_0}
	}_{= \lambda_0 \bra{\lambda_0}  \frac{\partial }{ \partial c_{j, k} }  \ket{\lambda_0} }
	-
	\lambda_0 \bra{\lambda_0}  \frac{\partial }{ \partial c_{j, k} }  \ket{\lambda_0}
}_{=0}
~,
\end{align}
which yields
\begin{align}
\frac{\partial \lambda_0 }{ \partial c_{j, k} }
= \braket{\lambda_0 | j} \braket{k | \lambda_0}
= \braket{k | \lambda_0} \braket{\lambda_0 | j} ~.
\end{align}
The summations over $\lambda_0$ thus become
\begin{align}
\sum_{\lambda_0 \in \Lambda_{\rho_0}}  \frac{\partial \lambda_0 }{ \partial c_{j, k} }
&= \sum_{\lambda_0 \in \Lambda_{\rho_0}} \braket{k | \lambda_0} \braket{\lambda_0 | j}
= \bra{ k } \Bigl( \underbrace{ \sum_{\lambda_0 \in \Lambda_{\rho_0}} \ket{\lambda_0} \bra{\lambda_0} }_{=I} \Bigr) \ket{ j }
= \braket{k | j}
= \tr \bigl( \ket{ j} \bra{ k} \bigr)
\label{eq:dr0sum}
\end{align}
and
\begin{align}
\sum_{\lambda_0 \in \Lambda_{\rho_0}} \Bigl( \frac{\partial \lambda_0 }{ \partial c_{j, k} }  \Bigr) \ln \lambda_0
&= \sum_{\lambda_0 \in \Lambda_{\rho_0}} \braket{k | \lambda_0} \braket{\lambda_0 | j} \ln \lambda_0
= \bra{ k } \Bigl( \underbrace{ \sum_{\lambda_0 \in \Lambda_{\rho_0}} \ln \lambda_0 \ket{\lambda_0} \bra{\lambda_0} }_{= \ln \rho_0} \Bigr) \ket{ j }
= \braket{k | \ln \rho_0 | j}
= \tr \bigl( \ket{ j} \bra{ k} \ln \rho_0  \bigr)
~.
\label{eq:dr0lnr0sum}
\end{align}

Moving on to the slightly more involved perturbation,
we use the eigen-relation:
$ \rho_\tau \ket{\lambda_\tau} = \lambda_\tau \ket{ \lambda_\tau } $
and again take the partial derivative of each side:
\begin{align}
\frac{\partial }{ \partial c_{j, k} }  \rho_\tau \ket{\lambda_\tau}
&= \Bigl( \frac{\partial }{ \partial c_{j, k} }  \rho_\tau \Bigr) \ket{\lambda_\tau}
+ \rho_\tau \frac{\partial }{ \partial c_{j, k} }  \ket{\lambda_\tau}
\qquad
=
\qquad
\frac{\partial }{ \partial c_{j, k} }  \lambda_\tau \ket{\lambda_\tau}
= \Bigl( \frac{\partial \lambda_\tau }{ \partial c_{j, k} }  \Bigr) \ket{\lambda_\tau}
+ \lambda_\tau \frac{\partial }{ \partial c_{j, k} }  \ket{\lambda_\tau}
~.
\end{align}
Left-multiplying by $\bra{\lambda_\tau}$,
and recalling that
$\rho_\tau = \sum_{j, k} c_{j, k} \trenv \bigl(
\mathcal{U}_{\drive}
\ket{j} \bra{k}
\otimes \rho_0^\text{env}
\mathcal{U}_{\drive}^\dagger
\bigr)$,
we obtain
\begin{align}
\Bigl( \frac{\partial \lambda_\tau }{ \partial c_{j, k} }  \Bigr) \braket{\lambda_\tau |\lambda_\tau}
=
\frac{\partial \lambda_\tau }{ \partial c_{j, k} }
=
\bra{\lambda_\tau} \Bigl( \!\!\!\!\! \!\!\!\!\! \underbrace{ \frac{\partial }{ \partial c_{j, k} }  \rho_\tau  }_{\trenv \bigl(
	\mathcal{U}_{\drive}
	\ket{j} \bra{k}
	\otimes \rho_0^\text{env}
	\mathcal{U}_{\drive}^\dagger
	\bigr)} \!\!\!\!\! \!\!\!\!\! \Bigr) \ket{\lambda_\tau}
+
\underbrace{
	\underbrace{
		\bra{\lambda_\tau} \rho_\tau \frac{\partial }{ \partial c_{j, k} }  \ket{\lambda_\tau}
	}_{= \lambda_\tau \bra{\lambda_\tau}  \frac{\partial }{ \partial c_{j, k} }  \ket{\lambda_\tau} }
	-
	\lambda_\tau \bra{\lambda_\tau}  \frac{\partial }{ \partial c_{j, k} }  \ket{\lambda_\tau}
}_{=0}
~,
\end{align}
which yields
\begin{align}
\frac{\partial \lambda_\tau }{ \partial c_{j, k} }
= \braket{\lambda_\tau | \trenv \bigl(
	\mathcal{U}_{\drive}
	\ket{j} \bra{k}
	\otimes \rho_0^\text{env}
	\mathcal{U}_{\drive}^\dagger
	\bigr) | \lambda_\tau}
~.
\end{align}
The summations over $\lambda_\tau$ thus become
\begin{align}
\sum_{\lambda_\tau \in \Lambda_{\rho_\tau}}  \frac{\partial \lambda_\tau }{ \partial c_{j, k} }
&= \sum_{\lambda_\tau \in \Lambda_{\rho_\tau}}
\braket{\lambda_\tau | \trenv \bigl(
	\mathcal{U}_{\drive}
	\ket{j} \bra{k}
	\otimes \rho_0^\text{env}
	\mathcal{U}_{\drive}^\dagger
	\bigr) | \lambda_\tau}
= \tr \Bigl( \trenv \bigl(
\mathcal{U}_{\drive}
\ket{j} \bra{k}
\otimes \rho_0^\text{env}
\mathcal{U}_{\drive}^\dagger
\bigr) \Bigr)
\\
&= \tr \bigl(
\mathcal{U}_{\drive}
\ket{j} \bra{k}
\otimes \rho_0^\text{env}
\mathcal{U}_{\drive}^\dagger
\bigr)
=   \tr \bigl(
\ket{j} \bra{k}
\otimes \rho_0^\text{env}
\bigr)
= \tr \bigl( \ket{ j} \bra{ k} \bigr)
\label{eq:drtausum}
\end{align}
and
\begin{align}
\sum_{\lambda_\tau \in \Lambda_{\rho_\tau}} \Bigl( \frac{\partial \lambda_\tau }{ \partial c_{j, k} }  \Bigr) \ln \lambda_\tau
&= \sum_{\lambda_\tau \in \Lambda_{\rho_\tau}} \braket{\lambda_\tau | \trenv \bigl(
	\mathcal{U}_{\drive}
	\ket{j} \bra{k}
	\otimes \rho_0^\text{env}
	\mathcal{U}_{\drive}^\dagger
	\bigr) | \lambda_\tau}  \ln \lambda_\tau
\\
&= \tr \biggl(  \Bigl( \underbrace{ \sum_{\lambda_\tau \in \Lambda_{\rho_\tau}} \ln \lambda_\tau \ket{\lambda_\tau} \bra{\lambda_\tau} }_{= \ln \rho_\tau} \Bigr)
\trenv \bigl(
\mathcal{U}_{\drive}
\ket{j} \bra{k}
\otimes \rho_0^\text{env}
\mathcal{U}_{\drive}^\dagger
\bigr)
\biggr)
\\
&=
\tr \Bigl(
\trenv \bigl(
\mathcal{U}_{\drive}
\ket{j} \bra{k}
\otimes \rho_0^\text{env}
\mathcal{U}_{\drive}^\dagger
\bigr)
\ln \rho_\tau
\Bigr)
~.
\label{eq:drtaulnrtausum}
\end{align}

Plugging in our new expressions
for the $\lambda_0$ an $\lambda_\tau$ summations in
Eqs.~\eqref{eq:dr0sum}, \eqref{eq:dr0lnr0sum}, \eqref{eq:drtausum}, and
\eqref{eq:drtaulnrtausum},
we obtain
\begin{align}
		\tfrac{1}{\kB} \frac{\partial}{ \partial c_{j, k} }  \EP
		&=
		\Bigl( \!\!\! \sum_{\lambda_0 \in \Lambda_{\rho_0}} \! \! \Bigl( \frac{\partial \lambda_0}{ \partial c_{j, k} } \Bigr) \ln \lambda_0 + \frac{\partial \lambda_0}{ \partial c_{j, k} }  \Bigr)
		-
		\Bigl( \!\!\! \sum_{\lambda_\tau \in \Lambda_{\rho_\tau}} \! \! \Bigl( \frac{\partial \lambda_\tau}{ \partial c_{j, k} } \Bigr) \ln \lambda_\tau + \frac{\partial \lambda_\tau}{ \partial c_{j, k} }  \Bigr)
		-
		\tr \bigl( \trsys ( \mathcal{U}_{\drive} \! \ket{j} \! \bra{k}
		\otimes \rho_0^\text{env}  \mathcal{U}_{\drive}^\dagger ) \ln  \stationary^\text{env} \bigr) \nonumber
		\\
		&=
		\tr \bigl( \ket{j} \bra{k} \ln \rho_0 \bigr)
		-
		\tr \Bigl( \trenv \bigl(
		\mathcal{U}_{\drive}
		\ket{j} \bra{k}
		\otimes \rho_0^\text{env}
		\mathcal{U}_{\drive}^\dagger
		\bigr)
		\ln \rho_\tau
		\Bigr)
		-
		\tr \Bigl( \trsys \bigl(
		\mathcal{U}_{\drive}
		\ket{j} \bra{k}
		\otimes \rho_0^\text{env}
		\mathcal{U}_{\drive}^\dagger
		\bigr)
		\ln \stationary^\text{env}
		\Bigr)
		~.
\label{eq:EPGradientElement}		
\end{align}

Recall that we have introduced a gradient
$\grad \EP \equiv \sum_{j, k} \ket{k} \bra{j} \frac{\partial}{\partial c_{j, k}} \EP$
and a scalar product
$\gamma \cdot \grad \EP
\equiv
\tr( \gamma \grad \EP )$.

This allows us to inspect local changes in entropy production
$(\rho_0 - \rho_0') \cdot \grad \EP[\rho_0']$
as we move from $\rho_0'$ towards any other density matrix $\rho_0$.
By the convexity of quantum states,
there is indeed a continuum of density matrices
in this direction; so
the sign of the directional derivative
indeed indicates the sign of the change in entropy production
for infinitesimal changes to the initial density matrix in the direction of $\rho_0$.

For any two density matrices $\rho_0 = \sum_{j,k} c_{j,k} \ket{j} \! \bra{k}$ and $\rho_0' = \sum_{j,k} c_{j,k}' \ket{j} \! \bra{k}$,
Eq.~\eqref{eq:EPGradientElement}
implies that
\begin{align}
	\tfrac{1}{\kB}
	\rho_0 \cdot \! \grad \EP[\rho_0']
	&=
	\tfrac{1}{\kB}
	\tr(\rho_0 \grad \EP[\rho_0']) \\
	&= \tfrac{1}{\kB} \sum_{j,k} c_{j,k}
	   \frac{\partial}{ \partial c_{j, k}' }  \EP[\rho_0'] \\
	&=
	\tr( \rho_0 \ln \rho_0' ) - \tr( \rho_\tau \ln \rho_\tau' )
		- \tr( \rho_\tau^\text{env} \ln \stationary^\text{env} )
		~.
\end{align}
so that we arrive at Lem.~\ref{lem:MismatchDot} of the main text.

\end{widetext}

\section{Useful Lemmas}
\label{sec:UsefulLemmas}

Here we derive useful lemmas about entropy production for arbitrary decompositions of the initial density matrix.
We consider any set of component density matrices $\{ \rho_0^{(n)} \}_n$ and corresponding probabilities $\{ \Pr(n) \}_n$
such that $\rho_0 = \sum_n \Pr(n) \rho_0^{(n)}$.

{\Lem \label{lem:ConvexDissipation}
Dissipation is convex over initial density matrices.
}

Let $\rho_0 = \sum_n \Pr(n) \rho_0^{(n)}$.
Then:

\vspace{-0.5em}

\scalebox{0.85}{
$\!\!\!\!\!\!\!\!$
\begin{minipage}{1.2\columnwidth}
\begin{align}
& \Bigl( \sum_n \Pr(n) \EP[\rho_0^{(n)}] \Bigr) - \EP \nonumber \\
& \; \, = \Bigl[ S(\rho_0) -  \sum_n \Pr(n) S(\rho_0^{(n)}) \Bigr]
    - \Bigl[ S(\rho_\tau) -  \sum_n \Pr(n) S(\rho_\tau^{(n)}) \Bigr]
\label{eq:Holevo}
    \\
& \; \, = \Bigl[  \sum_n \Pr(n) \text{D} \bigl[ \rho_0^{(n)} \big\| \rho_0 \bigr] \Bigr]
    - \Bigl[  \sum_n \Pr(n) \text{D} \bigl[ \rho_\tau^{(n)} \big\| \rho_\tau \bigr] \Bigr]
\label{eq:HolevoAsRelEnt}
    \\
& \; \, \geq 0 ~.
\label{eq:HolevoInfoProcessing}
\end{align}
\end{minipage}
}

\vspace{0.5em}

Eq.~\eqref{eq:Holevo} is obtained from the recognition that entropy flow $\EF$ is an affine function of $\rho_0$,
and so cancels between $\EP$
and $\sum_n \Pr(n) \EP[\rho_0^{(n)}] $.
Each of the square brackets then represents a Holevo information,
before and after the transformation, respectively.
The non-negativity is finally obtained by the information processing inequality applied to the Holevo information.
(Non-negativity can be seen to follow from each $n$ separately: I.e.,
$ \text{D} \bigl[ \rho_0^{(n)} \big\| \rho_0 \bigr]  \geq  \text{D} \bigl[ \rho_\tau^{(n)} \big\| \rho_\tau \bigr] $.)

{\Lem \label{lem:DissipationConvexEquality}
Dissipation is linearly combined as
$\EP =  \sum_n \Pr(n) \EP[\rho_0^{(n)}]$ if the elements of $\{ \rho_0^{(n)} \}_n$ are mutually orthogonal
and also the elements of $\{ \rho_\tau^{(n)} \}_n$ are mutually orthogonal.
}

If $\rho_t^{(m)}$ and  $\rho_t^{(n)}$ are orthogonal
(i.e., $\tr(\rho_t^{(m)} \rho_t^{(n)} ) = 0$ when $m \neq n$),
then they are simultaneously diagonalizable.
If this orthogonality holds for all pairs in the set $\{ \rho_t^{(n)} \}_n$,
then the simultaneous diagonalizability implies that
$\rho_t = \sum_n \Pr(n) \rho_t^{(n)}$ inherits the eigenvalues
of $\rho_t^{(n)}$ multiplied by their respective $\Pr(n)$.
I.e.:
$\rho_t$'s spectrum is then
\begin{align}
\Lambda_{\rho_t} = \bigcup_n \bigl\{ \lambda \Pr(n) : \lambda \in \Lambda_{\rho_0^{(n)}} \bigr\} ~,
\end{align}
with multiplicities inherited from the constituent spectra.

In this scenario, the von Neumann entropy cleaves into two pieces:
\begin{align}
S(\rho_t)
&= - \kB \sum_{\zeta \in \Lambda_{\rho_t}} \zeta \ln \zeta \\
&= - \kB \sum_n \sum_{\lambda \in \Lambda_{\rho_t^{(n)}}} \lambda \Pr(n) \ln \bigl(  \lambda \Pr(n) \bigr) \\
&= - \kB \Bigl( \sum_n \Pr(n) \sum_{\lambda \in \Lambda_{\rho_t^{(n)}}} \lambda  \ln \lambda \Bigr)
  \nonumber \\
  & \qquad  - \kB \sum_n \Pr(n) \ln \bigl( \Pr(n) \bigr)  \sum_{\lambda \in \Lambda_{\rho_t^{(n)}}} \lambda  \\
&=  \Bigl( \sum_n \Pr(n) S(\rho_0^{(n)})  \Bigr) - \kB \sum_n \Pr(n) \ln \bigl( \Pr(n) \bigr) ~,
\end{align}
where we used the fact that
$ \sum_{\lambda \in \Lambda_{\rho_t^{(n)}}} \lambda = 1$.
The \emph{difference} in entropy then yields

\vspace{-0.5em}

\scalebox{0.92}{
$\!\!\!\!\!\!$
\begin{minipage}{1.1\columnwidth}
\begin{align}
S(\rho_0) - S(\rho_\tau)
&=
\Bigl( \sum_n \Pr(n) S(\rho_0^{(n)})  \Bigr)
  - \Bigl( \sum_n \Pr(n) S(\rho_\tau^{(n)})  \Bigr) ~.
\end{align}
\end{minipage}
}

\vspace{0.5em}

Together with Eq.~\eqref{eq:Holevo}, this implies that
$\EP =  \sum_n \Pr(n) \EP[\rho_0^{(n)}]$,
proving Lemma \ref{lem:DissipationConvexEquality}.

\section{Generalized $\q_0$ for non-interacting basins}
\label{sec:Genq0}

It is possible that the
evolution acts completely independently on
distinct basins of state space.
This will generically yield a nontrivial nullspace
for $\argmin_{\rho_0} \EP$.
In such cases,
it is profitable to generalize the definition of $\q_0$
so that it includes the
successive minimally dissipative density matrices
that carve out the independent basins on the nullspace.
These basins will be formally defined shortly.

We use the terminology `basin' in analogy with `basins of attraction' in nonlinear dynamics.  Indeed, for certain dynamics that induce autonomous nonequilibrium steady states, this section addresses the quantum thermodynamics of coexisting basins of attraction for an open quantum system.
However, this section more generally addresses non-autonomous dynamics,
and allows for the possibility of thermodynamically independent regions of state space.  When a basin is dissipative, our results imply that trajectories from the initial conditions inside it must become less distinguishable over time---they contract to a generalized time-dependent notion of attractor.

We will show that, within each basin,
the extra dissipation due to a non-minimally dissipative initial density matrix
is given exactly by the contraction of the relative entropy between the actual and minimally dissipative
initial density matrices under the same driving $\drive$.

To make progress in this generalized setting,
we must first
introduce several new notions.

\subsection{Definitions}

Let $\mathcal{P}(\mathcal{H})$ be the set of density matrices that can be constructed
on the Hilbert space $\mathcal{H}$.
I.e.,
\scalebox{0.76}{
$\!\!\!\!$
\begin{minipage}{1.3\columnwidth}
\begin{align}
\mathcal{P}(\mathcal{H}) \equiv \Bigl\{ \sum_{\ell} \Pr(\ell) \frac{\ket{\ell} \bra{\ell}}{\braket{\ell | \ell}} : \ket{\ell} \in \mathcal{H} \! \setminus \! \{ \vec{0} \} , \, 0 \! < \! \Pr(\ell) \! \leq \! 1 ,  \text{ and } \sum_{\ell} \Pr(\ell) = 1  \Bigr\} .
\end{align}
\end{minipage}
}

Let
$\mathcal{H}^\text{sys}$ be the Hilbert space of the physical system under study (not including the environment).
We will denote the nullspace of an operator $\rho$ as null$(\rho)$.
I.e.:
$\text{null}(\rho) = \bigl\{ \ket{\eta} \in \mathcal{H}^\text{sys} : \rho \ket{\eta} = \vec{0} \bigr\}$.
The `support' of a density matrix is the space orthogonal to its nullspace.

We can now introduce the successive minimally dissipative density matrices $\{ \q_0^{[n]} \}_n$.
The absolute minimum dissipation is achieved via the minimally dissipative density matrix $\q_0^{[0]}$:
\begin{align}
\q_0^{[0]} \in \argmin_{\rho_0 \in \mathcal{P}(\mathcal{H}^\text{sys})}  \EP ~.
\end{align}
In the main text, where $\q_0^{[0]}$ has a trivial nullspace (of $\{ \vec{0} \}$),
we identified $\q_0^{[0]}$ with $\q_0$ itself.
However, when $\q_0^{[0]}$ has a nontrivial nullspace,
we will also want to consider the minimally dissipative density matrix on the nullspace:
$\q_0^{[1]} \in \argmin_{\rho_0 \in \mathcal{P} \bigl( \text{null}( \q_0^{[0]} ) \bigr)}  \EP$.
If $\q_0^{[1]}$ \emph{also} has a nontrivial nullspace, then we continue in the same fashion to identify
the minimally dissipative density matrix within the intersection of all of the preceding nullspaces.
In general, the $n^{\text{th}}$ thermodynamically independent basin has
the minimally dissipative initial state:
\begin{align}
\q_0^{[n]} \in \argmin_{\rho_0 \in \mathcal{P} \bigl( \bigcap_{m=0}^{n-1} \text{null}( \q_0^{[m]} ) \bigr)}  \EP
\end{align}
for $n \geq 1$.

The $n^\text{th}$ minimally dissipative basin
is the Hilbert space:
\begin{align}
\mathcal{H}_0^{[n]} = \text{span}\Bigl( \bigl\{ \ket{\lambda} : \lambda \in \Lambda_{\q_0^{[n]}} \setminus  0  \bigr\} \Bigr) ~,
\label{eq:BasinDef}
\end{align}
which is the support of $\q_0^{[n]}$.
We call this space a `basin'
in loose analogy with the basins of attraction of classical nonlinear dynamics.
We will employ the projector
\begin{align}
\Pi_{\mathcal{H}_0^{[n]}} = \sum_{\lambda \in \Lambda_{\! \q_0^{[n]}} \setminus \{ 0 \} } \frac{ \ket{\lambda} \bra{\lambda} }{ \braket{\lambda | \lambda} }
\end{align}
which projects onto $\mathcal{H}_0^{[n]}$.
Notably,
these projectors constitute a decomposition of the identity $I$ on the system's state space $\mathcal{H}^\text{sys}$:
\begin{align}
\sum_n \Pi_{\mathcal{H}_0^{[n]}} = I ~.
\label{eq:ProjectorDecomp}
\end{align}

We can now define the minimally dissipative reference state $\q_0$,
as if $\rho_0$ were minimally dissipative on each of the thermodynamically independent basins
on which it lives:
\begin{align}
\q_0 \equiv \sum_n \tr( \Pi_{\mathcal{H}_0^{[n]} } \rho_0 ) \q_0^{[n]} ~.
\label{eq:gen_q0_def}
\end{align}
It should be noted that the $\rho_0$-dependence is only via
the weight $\tr(\Pi_{\mathcal{H}_0^{[n]} } \rho_0)$ of
$\rho_0$ on each thermodynamically-independent basin,
used to linearly combine their contributions.

\subsection{Generalized dissipation bound}

With these definitions in place,
let us now reconsider the task at hand.

If $\q_0^{[0]}$
has a nontrivial nullspace
(and if $\EP$ is finite for all $\rho_0$),
then there are
thermodynamically isolated
basins of state-space.
(The support of $\q_0^{[0]}$ is the minimally dissipative basin $\mathcal{H}_0^{[n]}$, which is a strict subset of $\mathcal{H}^\text{sys}$.)
Then,
since a generic initial state $\rho_0$ may have support on the nullspace of $\q_0^{[0]}$,
Eq.~\eqref{eq:ExtremumProperty} is no longer directly valid.
However,
for any two initial  density matrices $\rho_0$ and $\xi_0$
such that the
support of $\rho_0$ is a subset of the
support of $\xi_0$,
it is still true that
\begin{align}
& (\rho_0 - \xi_0) \cdot \grad \EP[\xi_0]
\nonumber \\
& \quad = \EP - \EP[\xi_0]
  - \kB \Bigl(  \text{D} \bigl[ \rho_0 \big\| \xi_0 \bigr] - \text{D} \bigl[ \rho_\tau \big\| \xi_\tau \bigr]   \Bigr)  ~.
\label{eq:Lemma1Implication}
\end{align}
Indeed $\q_0$, as defined in Eq.~\eqref{eq:gen_q0_def},
is guaranteed to have
support equal to
$\mathcal{H}^\text{sys}$, and so Eq.~\eqref{eq:Lemma1Implication}
is valid if we set $\xi_0 = \q_0$.
Alternatively, we can set $\xi_0 = \q_0^{[n]}$ if we properly restrict $\rho_0$.

To proceed, we recognize that $\rho_0$ can be
decomposed
via Eq.~\eqref{eq:ProjectorDecomp}
as
\begin{align}
\rho_0
&=
\rho_0^\Box
  + \rho_0^\text{coh}
~,
\label{eq:BasinAndCoherenceDecomp}
\end{align}
where
\begin{align}
\rho_0^\Box &\equiv
\sum_n \tr( \Pi_{\mathcal{H}_0^{[n]}} \rho_0 ) \rho_0^{[n]}
& \, \text{ with } \,
\rho_0^{[n]} \equiv \frac{  \Pi_{\mathcal{H}_0^{[n]}} \rho_0  \Pi_{\mathcal{H}_0^{[n]}}  }{  \tr( \Pi_{\mathcal{H}_0^{[n]}} \rho_0 )  }
\nonumber
\intertext{and}
\rho_0^\text{coh} &\equiv  \sum_{m, n \atop m \neq n}   \Pi_{\mathcal{H}_0^{[m]}} \rho_0  \Pi_{\mathcal{H}_0^{[n]}} ~.
\nonumber
\end{align}
$\rho_0^\Box$ projects $\rho_0$ onto the minimally dissipative basins,
whereas
$\rho_0^\text{coh} $
describes the state's coherence \emph{between} these basins.

Since $\q_0^{[n]}$ is, by definition,
the minimally dissipative density matrix on its subspace
(and, since it has full support on that subspace),
we have that
\begin{align}
0
&=
( \rho_0^{[n]}   - \q_0^{[n]} )
 \cdot \grad \EP[\q_0^{[n]} ]
\\
& = \EP[\rho_0^{[n]}] - \EP[\q_0^{[n]}]
  - \kB \Bigl(  \text{D} \bigl[ \rho_0^{[n]} \big\| \q_0^{[n]} \bigr] - \text{D} \bigl[ \rho_\tau^{[n]} \big\| \q_\tau^{[n]} \bigr]   \Bigr) ~.
\label{eq:MainEqForEach_n}
\end{align}

As an immediate consequence of their definition, the elements
of $\{ \q_0^{[n]} \}_n$ are mutually orthogonal,
and
the elements
of $\{ \rho_0^{[n]} \}_n$ are mutually orthogonal.
It is often the case (and, we conjecture, generally true)
that the coexistence of these $\q_0^{[n]}$ basins
implies the orthogonality of their evolved states.
I.e., the elements of $\{ \q_\tau^{[n]} \}_n$ are mutually orthogonal
and the elements of $\{ \rho_\tau^{[n]} \}_n$ are mutually orthogonal.
Lemma \ref{lem:DissipationConvexEquality} then implies that
\begin{align}
\EP[\q_0] =  \sum_n \tr( \Pi_{\mathcal{H}_0^{[n]} } \rho_0 ) \EP[\q_0^{[n]}]
\end{align}
and
\begin{align}
\EP[\rho_0^\Box] =  \sum_n \tr( \Pi_{\mathcal{H}_0^{[n]} } \rho_0 ) \EP[\rho_0^{[n]}]  ~.
\end{align}
Together with Eq.~\eqref{eq:MainEqForEach_n}
this leads to
\begin{align}
\EP[\rho_0^\Box] - \EP[\q_0]
&=
  \kB \Bigl(  \text{D} \bigl[ \rho_0^\Box \big\| \q_0 \bigr] - \text{D} \bigl[ \rho_\tau^\Box \big\| \q_\tau \bigr]   \Bigr) ~.
\label{eq:MainEq_with_no_interblock_coherence}
\end{align}

Eq.~\eqref{eq:MainEq_with_no_interblock_coherence}
can be seen as the quantum generalization of
the classical result obtained recently in Ref.~\cite{Wolp19}.
The classical version of this result is relevant when
the minimally dissipative probability distribution ($\q_0^{[0]}$)
does not have full support.
Dissipation on other `islands' are then considered.
Our derivation points out the nuances of physical assumptions that go into
the classical result, and refines the notion of `islands' (here referred to as `basins')
on a more solid physical grounding.
Crucially,
Eq.~\eqref{eq:MainEq_with_no_interblock_coherence}
generalizes the classical result---allowing $\rho_0^{[n]}$
to exhibit quantum
coherence relative to the minimally dissipative state $\q_0^{[n]}$.
In addition to the drop in Kullback--Leibler divergence on the minimally dissipative eigenbasis,
the drop in coherence also contributes to dissipation.

In the quantum regime, there is yet further opportunity for generalization,
if we consider the possibility of coherence among the non-interacting basins of state-space.
This is the case of non-zero inter-basin coherence: $\rho_0^\text{coh} \neq 0$.
To address this more general case,
we recognize that
\begin{align}
\EP - \EP[\rho_0^\Box]
  =   \EF^\text{coh}
      - \kB \Delta C_{\rho_t^\Box} (\rho_t) ~,
\label{eq:BoxCorrection_toEntropy}
\end{align}
where $\Delta C_{\rho_t^\Box} (\rho_t) $ is the change in
\emph{inter-basin coherence}:
\begin{align}
C_{\rho_t^\Box} (\rho_t) \equiv
\tr(\rho_t \ln \rho_t)
  -  \tr(\rho_t^\Box \ln \rho_t^\Box)
~,
\end{align}
from time $t=0$ to time $t=\tau$.
The inter-basin coherence $C_{\rho_t^\Box} (\rho_t)$
can also be recognized as the ``relative entropy of superposition''
among the basins~\cite{Aber06}.
Meanwhile,
\begin{align}
\EF^\text{coh}
&=
- \kB \tr \bigl( \trsys ( \mathcal{U}_{\drive}  \rho_0^\text{coh} \otimes \rho_0^\text{env}  \mathcal{U}_{\drive}^\dagger ) \ln  \stationary^\text{env} \bigr)
\end{align}
is the extra
entropy flow due to inter-basin coherence.

Combining Eqs.~\eqref{eq:MainEq_with_no_interblock_coherence} and \eqref{eq:BoxCorrection_toEntropy}
yields
\begin{align}
\EP - \EP[\q_0]
&=
  \kB \Bigl(  \text{D} \bigl[ \rho_0 \big\| \q_0 \bigr] - \text{D} \bigl[ \rho_\tau \big\| \q_\tau \bigr]   \Bigr)
  +
  \EF^\text{coh}
    ~.
\label{eq:MainEq_with_interblock_coherence}
\end{align}

All results of this section can be directly translated to
the generalized quantum optimization 
problem discussed in Sec.~\ref{sec:OtherOptProblems},
by simply replacing $\EP[\rho] / \kB$ with $f(\rho)$,
and replacing $\EF^\text{coh} / \kB$ with $\ell(\rho_0^\text{coh})$.

\section{Unitary Subspaces of Nonunitary Transformations}
\label{sec:UnitarySubspaces}

If the quantum operation $\QProcess(\cdot)$ is locally unitary on its restriction to
$\mathcal{P}(\mathcal{H}_0^{[n]})$
(i.e., restricted to density operators defined on the support of $\q_0^{[n]}$),
then any $\rho_0^{[n]} \in \mathcal{P}(\mathcal{H}_0^{[n]})$ has the same dissipation
(i.e., $\EP[\rho_0^{[n]}] = \EP[\q_0^{[n]}]$).

Conversely,
if  $\QProcess(\cdot)$ has no nontrivial unitary subspaces on its restriction to
$\mathcal{P}(\mathcal{H}_0^{[n]})$,
then
any $\rho_0^{[n]} \neq \q_0^{[n]}$
\emph{requires} additional
dissipation.
Whenever the Second Law is valid,
this is additional dissipation beyond Landauer's bound.
In particular,
from Eq.~\eqref{eq:MainEqForEach_n}:
\begin{align}
\EP[\rho_0^{[n]}] = \EP[\q_0^{[n]}] + \kB
\Bigl(  \text{D} \bigl[ \rho_0^{[n]} \big\| \q_0^{[n]} \bigr] - \text{D} \bigl[ \rho_\tau^{[n]} \big\| \q_\tau^{[n]} \bigr]   \Bigr) ~.
\end{align}
Since we are assuming that the quantum operation $\QProcess(\cdot)$ is strictly nonunitary on this subspace,
$ \text{D} \bigl[ \rho_0^{[n]} \big\| \q_0^{[n]} \bigr] - \text{D} \bigl[ \QProcess(\rho_0^{[n]}) \big\| \QProcess(\q_0^{[n]}) \bigr]  $
is necessarily positive for $\rho_0^{[n]} \neq \q_0^{[n]}$.

Notably, an operation that is \emph{logically irreversible} on the computational basis (or, indeed, on any basis)
is always nonunitary and so cannot be thermodynamically optimized for all initial states via a single protocol.
Nonunitarity is more general than logical irreversibility though.

All quantum operations that are subject to either a Landauer-type cost or benefit
(due to the entropy change of nonunitary operations)
also incur irreversible dissipation for at least some initial states
(due to the relative entropy change of nonunitary operations).

\section{Change in relative entropy decomposes into change in D$_\text{KL}$s and change in coherences}
\label{sec:RelEntDecomp}

Our main result, Eq.~\eqref{eq:MainResult}, gave the extra dissipation---when the system
starts with the initial density matrix $\rho_0$ rather than the minimally-dissipative
initial density matrix $\q_0$---in terms of the change in relative entropies between the two reduced density matrices over the course of the transformation:
\begin{align}
\EP - \EP[\q_0] = \kB \text{D} \bigl[ \rho_0 \big\| \q_0 \bigr] - \kB \text{D} \bigl[ \rho_\tau \big\| \q_\tau \bigr] ~.
\label{eq:MainResultApp}
\end{align}
We now show how this can be split into a change in Kullback--Leibler divergences
plus the change in the coherence on the minimally-dissipative eigenbasis.

The correspondence with Ref.~\cite{Kolc17} is
complicated by the fact that $\q_0$ and $\q_\tau$
are not typically diagonalized in the same basis.
Nevertheless, we can consider $\q_t$'s eigenbasis at each time,
and describe $\rho_t$'s
probabilities and coherence relative to that basis.

The classical probability distribution
that would be induced by projecting $\rho_t$ onto $\q_t$'s eigenbasis is $\mathcal{P}_t$,
which can be represented as
\begin{align}
P_t \equiv
\sum_{s \in \Lambda_{\q_t}} \mathcal{P}_t(s) \ket{s} \bra{s}
~,
\end{align}
where the probability elements are $\mathcal{P}_t(s) = \braket{s | \rho_t | s}$.
The operators
$\rho_t$ and $P_t$ only differ when $\rho_t$ is coherent on $\q_t$'s eigenbasis.

The actual state's coherence on $\q_t$'s eigenbasis
is given by the `relative entropy of coherence'~\cite{Baum14}:
\begin{align}
C_{\q_t}(\rho_t)
&=
\tr(\rho_t \ln \rho_t) - \tr(P_t \ln P_t)
~.
\end{align}

Expanding the relative entropy between $\rho_t$ and $\q_t$ at any time yields
\begin{align}
\text{D} \bigl[ \rho_t \big\| \q_t \bigr]
&= \tr(\rho_t \ln \rho_t) - \tr(\rho_t \ln \q_t) \\
&= C_{\q_t}(\rho_t)
+ \tr(P_t \ln P_t)
- \tr(\rho_t \ln \q_t)
\\
&=
C_{\q_t} \! (\rho_t) \! + \!\!\!
\sum_{s \in \Lambda_{\q_t}}  \!\!\! \mathcal{P}_t(s) \ln \bigl[ \mathcal{P}_t(s) \bigr] \! - \! \mathcal{P}_t(s) \ln \bigl[ \mathcal{Q}_t(s) \bigr]
\\
&=
C_{\q_t}(\rho_t) + \text{D}_\text{KL} \bigl[ \mathcal{P}_t \big\| \mathcal{Q}_t \bigr]
~,
\end{align}
where we used the simultaneously-diagonalized
spectral representations of $\ln P_t = \sum_{s \in \Lambda_{\q_t}} \ln \bigl[ \mathcal{P}_t(s) \bigr] \ket{s} \bra{s}$ and
$\ln \q_t = \sum_{s \in \Lambda_{\q_t}} \ln \bigl[ \mathcal{Q}_t(s) \bigr] \ket{s} \bra{s}$,
and where $\mathcal{P}_t(s) \equiv \bra{s} P_t \ket{s} = \bra{s} \rho_t \ket{s}$
and $\mathcal{Q}_t(s) \equiv \bra{s} \q_t \ket{s}$ are the probability elements of
the classical probability distributions $\mathcal{P}_t$ and $\mathcal{Q}_t$ on the simplex defined by
$\q_t$'s eigenstates.

(Similar decompositions of the quantum relative entropy appear in recent
thermodynamic results of Refs.~\cite{Fran19a} and \cite{Sant19a}, although in a more limited context.)

Thus, the difference in entropy production can be expressed as
\begin{align}
\tfrac{1}{\kB} ( \EP - \EP[\q_0] )
&=
\text{D} \bigl[ \rho_0 \big\| \q_0 \bigr] - \text{D} \bigl[ \rho_\tau \big\| \q_\tau \bigr]
\\
&=
\text{D}_\text{KL} \bigl[ \mathcal{P}_0 \big\| \mathcal{Q}_0 \bigr] -  \text{D}_\text{KL} \bigl[ \mathcal{P}_\tau \big\| \mathcal{Q}_\tau \bigr]
\nonumber \\
& \qquad \! + \,
C_{\q_0}(\rho_0) - C_{\q_\tau}(\rho_\tau)
~,
\label{eq:EPwithCoherences}
\end{align}
as in Eq.~\eqref{eq:EPasWKplusCoherence}
of the main text.

In the classical limit, where there are no coherences,
we recover the classical result obtained by Kolchinsky and Wolpert in Ref.~\cite{Kolc17}:
\begin{align}
\tfrac{1}{\kB} ( \EP - \EP[\q_0] )^\text{classical}
&=
\text{D}_\text{KL} \bigl[ \mathcal{P}_0 \big\| \mathcal{Q}_0 \bigr] -  \text{D}_\text{KL} \bigl[ \mathcal{P}_\tau \big\| \mathcal{Q}_\tau \bigr]
~.
\end{align}
From Eq.~\eqref{eq:EPwithCoherences},
we see that the quantum correction to the classical dissipation
is exactly the
change of
coherence on the minimally-dissipative eigenbasis.

\section{Justification for Approach to the Gibbs State under Weak Coupling}

Consider a system in constant
energetic contact with a single thermal bath
of inverse temperature $\beta = \tfrac{1}{\kB T}$.
Suppose the system experiences
a time-independent Hamiltonian $H_x$  (i.e., $x_t = x_{t'} = x$ for all $t, t' > 0$).

From Ref.~\cite{Lena78}, we can deduce that the system together with part of the thermal bath will together approach a stable passive state under the influence of the remainder of the thermal bath.  For large baths,
this stable passive state limits to the Gibbs state
for the joint system.
If we furthermore
take the limit of very weak coupling,
then this also yields the Gibbs state for the reduced system since
$e^{\beta (H_x \otimes I_b + I_\text{sys} \otimes H_b )} = e^{\beta H_x } \otimes e^{\beta H_b }$.
The system--bath interaction $H_x^\text{int}$ can be treated as a small perturbation to the steady state
with vanishing contribution in the limit of very weak coupling.

Hence, if this system starts out of equilibrium in state $\rho_0$,
then it will simply relax towards
the canonical equilibrium state
$\stationary_x = e^{- \beta H_x}/Z$,
where $Z$ is the canonical partition function of the system.

Similar reasoning justifies the approach to equilibrium in any thermodynamic potential.

The case of strong coupling is more tricky because of the possibility of steady-state coherences in the system's energy eigenbasis~\cite{Guar18a}.
Nevertheless, there are small quantum systems of significant interest that
are rigorously shown to approach the Gibbs state as an attractor, even with strong interactions~\cite{Gave17, Gave19}.

\section{Dissipation, Work, and Free Energy}

Time-dependent control implies work and,
in the thermodynamics of computation, entropy production is typically
proportional to
the \emph{dissipated work}~\cite{Parr15a, Riec19b}.
This appendix relates these quantities.

Since we allow for arbitrarily strong interactions between system and baths,
some familiar thermodynamic equations must be revised in recognition of
interaction energies.  Most of these revisions have already been thought through carefully
in Ref.~\cite{Espo10a}.
In this appendix, we spell out some of the general relationships among
entropy production, heat, work, dissipated work, nonequilibrium free energy, and so on.
This allows our results to be reinterpreted in terms of
the various thermodynamic quantities.

Work is the amount of energy pumped into the
system and baths
by the time-varying Hamiltonian.
It is the total change in energy of the system and baths:
\begin{align}
W \equiv \tr(\rho_\tau^\text{tot} H_{x_\tau}^\text{tot}) -  \tr(\rho_0^\text{tot} H_{x_0}^\text{tot}) ~.
\end{align}

Subtracting the heat yields
\begin{align}
W - \sum_{b \in \mathcal{B}} Q^{(b)}
&=
\tr(\rho_\tau H_{x_\tau}) -  \tr(\rho_0 H_{x_0})
\nonumber \\
& \quad +
\tr(\rho_\tau^\text{tot} H_{x_\tau}^\text{int}) -  \tr(\rho_0^\text{tot} H_{x_0}^\text{int})
 ~,
\end{align}
which generalizes the First Law of Thermodynamics
(as earlier noted in Ref.~\cite{Espo10a})
beyond the weak-coupling limit.
The typical First Law is recovered when the interaction energy is the same at the beginning and end of the protocol.
Alternatively, the First Law can be approximately achieved if the interaction energy is relatively weak at the beginning and end of the protocol.

If there is a single canonical bath at temperature $T$,
then entropy production is related to the dissipated work and the nonequilibrium free energy.
In that case, the \emph{dissipated work} is
\begin{align}
\Wdiss
&= T \EP
\\
&= Q +  \kB T  \tr ( \rho_0 \ln \rho_0 ) -  \kB T  \tr ( \rho_\tau \ln \rho_\tau )
\\
&= W -  (\mathcal{F}_\tau - \mathcal{F}_0) - \bigl( \tr(\rho_\tau^\text{tot} H_{x_\tau}^\text{int}) -  \tr(\rho_0^\text{tot} H_{x_0}^\text{int}) \bigr)
~.
\end{align}
We see that the dissipated work is the work beyond the changes
in nonequilibrium free energy and interaction energy.
Any work not stored in free energy or interaction energy has been dissipated.

In the presence of a single canonical bath,
the nonequilibrium free energy always satisfies the familiar relation $\mathcal{F}_t = U_t - TS_t$:
\begin{align}
\mathcal{F}_t
&\equiv
  \tr(\rho_t H_{x_t}) + \kB T \tr ( \rho_t \ln \rho_t ) \\
&=
F_{x_t}^\text{eq} + \mathcal{F}_t^\text{add}  ~,
\end{align}
where the nonequilibrium addition to free energy is
\begin{align}
 \mathcal{F}_t^\text{add}
 = \kB T \text{D} \bigl[ \rho_t \big\|  \stationary_{x_t} \bigr]  ~,
\end{align}
$ \stationary_{x_t} = e^{-\beta H_{x_t }} / Z_{x_t}$ is the Gibbs state induced by the instantaneous control,
and
$F_{x_t}^\text{eq} = - \kB T \ln Z_{x_t}$ is the equilibrium free energy of the system,
which utilizes the partition function
$Z_{x_t} = \tr( e^{-\beta H_{x_t }}  )$.

Even if the interaction energy is large, we see that we recover the familiar thermodynamic relations,
as long as there is negligible net change in interaction energy over the course of the protocol:
$\tr(\rho_\tau^\text{tot} H_{x_\tau}^\text{int}) -  \tr(\rho_0^\text{tot} H_{x_0}^\text{int}) \approx 0$.
Then,
$W - \sum_{b \in \mathcal{B}} Q^{(b)}
=
\tr(\rho_\tau H_{x_\tau}) -  \tr(\rho_0 H_{x_0}) $
and
$\Wdiss = W -  (\mathcal{F}_\tau - \mathcal{F}_0)$.

\section{High-fidelity \texttt{RESET} bounds relative entropy}
\label{sec:ResetBounds}

For any process whatsoever,
$0 \leq \text{D} \bigl[ \rho_\tau \big\| \q_\tau \bigr] \leq \text{D} \bigl[ \rho_0 \big\| \q_0 \bigr] $.
But we can also derive a number of stricter bounds.
Here we will show that
$\text{D} \bigl[ \rho_\tau \big\| \q_\tau \bigr]
\leq  4\epsilon \ln \Bigl( \frac{d}{4 \epsilon \sqrt{s_\text{min}}} \Bigr) $,
where $d$ is the dimension of the Hilbert space and
$s_\text{min}$ is the smallest eigenvalue of $\q_\tau$, and
$\epsilon$ upper bounds the trace distance between the desired state $r_\tau$
and the actual final state $\rho_\tau$ from any initial state $\rho_0$.

For any fixed dimension and $s_\text{min} \neq 0$,
the $\epsilon \to 0$ limit of high-fidelity \texttt{RESET} then yields
$\text{D} \bigl[ \rho_\tau \big\| \q_\tau \bigr] \to 0$,
which immediately leads to Eq.~\eqref{eq:ResetDiss}.

When $\epsilon$ and $s_\text{min}$ \emph{both} go to zero,
the situation is more delicate:
we obtain Eq.~\eqref{eq:ResetDiss}
only in the limit that $- \epsilon \ln( s_\text{min} ) \to 0$.
For example, this limit can be obtained if $\epsilon$ is never larger than $s_\text{min}$,
in which case $0 \leq -\epsilon \ln (s_\text{min}) \leq -\epsilon \ln (\epsilon) \to 0$ as $\epsilon \to 0$.

In either case:
$\text{D} \bigl[ \rho_\tau \big\| \q_\tau \bigr] \to 0$ as $\epsilon \to 0$ if $\epsilon \leq s_\text{min}$.

\subsection{Deriving the bound}

Recall that the trace distance $\mathcal{T}(\rho, \sigma)$
between two quantum states $\rho$ and $\sigma$ is given by:
$\mathcal{T}(\rho, \sigma) \equiv \tfrac{1}{2} \| \rho - \sigma \|_1$,
where $\| A \|_1 \equiv \tr(\sqrt{ A^\dagger A }) $ is the trace norm.
By definition, the error tolerance $\epsilon$ associated with a control protocol
implementing the \texttt{RESET} operation
is defined such that
$\mathcal{T}(\rho_\tau, r_\tau) \leq \epsilon$ for all $\rho_0$.

As an aside,
recall that the fidelity $F(\rho, \sigma) \equiv \tr \bigl( \sqrt{ \sqrt{\rho} \sigma \sqrt{\rho} } \bigr)$ and the trace distance $\mathcal{T}(\rho, \sigma)$
are related by
$1 - F(\rho, \sigma) \leq \mathcal{T}(\rho, \sigma) \leq \sqrt{1 - F(\rho, \sigma)^2 } $.
~\cite{Niel10a}~\footnote{We adopt the definition used in Refs.~\cite{Aude05, Niel10a}.  However, the fidelity is sometimes defined as the square of this.}  Accordingly:
\begin{align}
1 - F(\rho_\tau, r_\tau) \leq \mathcal{T}(\rho_\tau, r_\tau) \leq \epsilon ~,
\end{align}
which can be rearranged as
$F(\rho_\tau, r_\tau) \geq 1- \epsilon$.
So the limit of $\epsilon \to 0$ indeed corresponds to the limit of high fidelity $F(\rho_\tau, r_\tau) \to 1$.

Ref.~\cite[Thm.~3]{Aude05} provides a general bound on relative entropy between two density matrices which,
when applied to $\rho_\tau$ and $\q_\tau$, says:
\begin{align}
\text{D} \bigl[ \rho_\tau \big\| \q_\tau \bigr]
\leq
  \| \rho_\tau - \q_\tau \|_1 & \ln d
  - \| \rho_\tau - \q_\tau \|_1 \ln \| \rho_\tau - \q_\tau \|_1
  \nonumber \\
  &
  - \tfrac{1}{2} \| \rho_\tau - \q_\tau \|_1 \ln s_\text{min} ~,
\label{eq:AudenaertBound}
\end{align}
where $s_\text{min} \equiv \min \Lambda_{\q_\tau}$ is the smallest of $\q_\tau$'s eigenvalues
and $d$ is the dimension of the Hilbert space.
By definition, the error tolerance bounds the distance from the desired state to any final state; so
$\tfrac{1}{2} \| \rho_\tau - r_\tau \|_1 \leq \epsilon$ and $\tfrac{1}{2} \| \q_\tau - r_\tau \|_1 \leq \epsilon$.
From the triangle inequality, this implies that
\begin{align}
\| \rho_\tau - \q_\tau \|_1 \leq 4 \epsilon ~.
\end{align}
Applying this to Eq.~\eqref{eq:AudenaertBound}
yields a bound in terms of error tolerance:
\begin{align}
\text{D} \bigl[ \rho_\tau \big\| \q_\tau \bigr]
& \leq
  4 \epsilon \ln d
  - 4 \epsilon  \ln (4 \epsilon )
  - 2 \epsilon  \ln s_\text{min}  \\
& =
  4 \epsilon \ln \Bigl( \frac{d}{ 4 \epsilon \sqrt{ s_\text{min}}} \Bigr)
  ~.
\label{eq:ModifiedAudenaertBound}
\end{align}
Together with Eq.~\eqref{eq:MainResult},
this in turn implies a sandwiching of the entropy production:  \begin{align}
\text{D} \bigl[ \rho_0 \big\| \q_0 \bigr] - 4 \epsilon \ln \Bigl( \frac{d}{4 \epsilon \sqrt{ s_\text{min}}} \Bigr) \leq
\frac{ \EP - \EP[\q_0] }{\kB} \leq  \text{D} \bigl[ \rho_0 \big\| \q_0 \bigr] ~.
\end{align}
When $\epsilon \to 0$
and $\epsilon \ln (s_\text{min}) \to 0$,
the lower bound converges to the upper bound and we obtain
$\EP \to \EP[\q_0] + \kB \text{D} \bigl[ \rho_0 \big\| \q_0 \bigr] $.

Ref.~\cite{Aude05} also provides sharper bounds on relative entropy that could likely be leveraged
to give sharper bounds on the dissipation.  In particular, stronger bounds in Ref.~\cite{Aude05}
are independent of dimension and suggest that
$\text{D} \bigl[ \rho_\tau \big\| \q_\tau \bigr] \to 0$
much more quickly than suggested by Eq.~\eqref{eq:ModifiedAudenaertBound}
as $\epsilon \to 0$.

\begin{widetext}
	
\section{Modularity dissipation}
\label{sec:ModDiss}

This appendix gives more details of the derivation for the general modularity dissipation.

Consider a collection of elementary quantum operations $G_n$, acting on respective Hilbert space $\mathcal{H}_n$. Each is individually optimized,
such that dissipation is minimized for some $\q_{0,n}$. Suppose we place them in parallel to build a composite $N$-partite operation $\Gamma = \bigotimes_{n=1}^N G_n$.
Individual optimization implies that the minimally dissipative state, $\q_0 = \bigotimes_{n=1}^N \q_{0, n}$,
will take a product form. 
Likewise, the time-evolved 
minimally dissipative state
will retain this product structure:
$\q_\tau = \bigotimes_{n=1}^N \q_{\tau, n}$
where $\q_{\tau, n} = G_n(\q_{0, n})$.

We now consider a generic input $\rho_0$ to our computation.
Each elementary operation $G_n$ acts on a reduced state $\rho_{0,n}$ 
where $\rho_{t,n} = \tr_{\otimes_{m \neq n} \mathcal{H}_m} (\rho_t)$.
With our main result and a little algebra,
we calculate
\begin{align}
\tfrac{1}{\kB} \bigl( \EP - \EP[\otimes_n \rho_{0,n}] \bigr)
&=
\tfrac{1}{\kB} \big[( \EP - \EP[\q_0]) - (\EP[\otimes_n \rho_{0,n}] - \EP[\q_0]) \bigr] \\
&=
- \Delta \text{D} \bigl[ \rho_t \big\| \bigotimes_{n=1}^N \q_{t, n } \bigr]
+ \Delta \text{D} \bigl[ \bigotimes_{n=1}^N \rho_{t, n } \big\| \bigotimes_{n=1}^N \q_{t, n } \bigr] \\
&=
- \Delta \tr(\rho_t \ln \rho_t) + \Bigl[ \sum_{n=1}^N \Delta \tr(\rho_{t,n} \ln \q_{t, n }) \Bigr]
+ \Bigl[ \sum_{n=1}^N \Delta \tr(\rho_{t,n} \ln \rho_{t, n }) \Bigr]
- \Bigl[ \sum_{n=1}^N \Delta \tr(\rho_{t,n} \ln \q_{t, n }) \Bigr] \\
&=
- \Delta \tr(\rho_t \ln \rho_t) 
+ \Bigl[ \sum_{n=1}^N \Delta \tr(\rho_{t,n} \ln \rho_{t, n }) \Bigr] \\
&=
- \Delta \text{D} \bigl[ \rho_t \big\| \bigotimes_{n=1}^N \rho_{t, n } \bigr] ~.
\label{eq:ModularityDiss1app}
\end{align}
Eq.~\eqref{eq:ModularityDiss1app}
gives the modularity dissipation due to lost correlations in any finite-time composite quantum operation.
The right-hand side of Eq.~\eqref{eq:ModularityDiss1app}
is the reduction in \emph{total correlation} 
among the $N$ subsystems.

Furthermore, it may be noted that 
\begin{align}
\tfrac{1}{\kB} \bigl( \EP[\otimes_n \rho_{0,n}] - \EP[\q_0] \bigr)
&=
- \Delta \text{D} \bigl[ \bigotimes_{n=1}^N \rho_{t, n } \big\| \bigotimes_{n=1}^N \q_{t, n } \bigr] \\
&=
- \Bigl[ \sum_{n=1}^N \Delta \tr(\rho_{t,n} \ln \rho_{t, n }) \Bigr]
+ \Bigl[ \sum_{n=1}^N \Delta \tr(\rho_{t,n} \ln \q_{t, n }) \Bigr] \\
&=
- \sum_{n=1}^N \Delta \text{D} \bigl[ \rho_{t, n } \big\|  \q_{t, n } \bigr] ~.
\label{eq:ModularityDiss2app}
\end{align}

The total dissipation from a composite transformation can thus be written as:
\begin{align}
\EP 
&=
\EP[\q_0] 
- \kB \Delta \text{D} \bigl[ \rho_t \big\| \bigotimes_{n=1}^N \rho_{t, n } \bigr]
- \kB \sum_{n=1}^N \Delta \text{D} \bigl[ \rho_{t, n } \big\|  \q_{t, n } \bigr] ~.
\label{eq:ModularityDiss3app}
\end{align}
This is the sum of 
1)
modularity dissipation
$- \kB \Delta \text{D} \bigl[ \rho_t \big\| \bigotimes_{n=1}^N \rho_{t, n } \bigr]$
due to the loss of (both quantum and classical) correlation among subsystems,
2)
the mismatch dissipation $\kB \Delta \text{D} \bigl[ \rho_{t, n } \big\|  \q_{t, n } \bigr]$
from non-optimal input to each elementary operation,
and 3)
the residual dissipation $\EP[\q_0]$
which is invariably incurred by even the minimally dissipative input to the composite operation.

In the classical limit,
the  total correlation reduces to the Kullback--Leibler divergence
$\text{D} \bigl[ \rho_t \big\| \bigotimes_{n=1}^N \rho_{t, n } \bigr] \to \text{D}_\text{KL} \bigl[ \Pr(X_{t,1}, X_{t,2}, \dots, X_{t,N}) \big\| \prod_{n=1}^N \Pr(X_{t,n}) \bigr]$,
where $X_{t,n}$ is the random variable for the state of the $n^\text{th}$ subsystem at time $t$.
This then generalizes the modularity dissipation expected for classical computations, discussed in Refs.~\cite{Boyd18b, Riec19b}.
Notably, modularity dissipation 
can be written in the exact forms of either 
Eqs.~\eqref{eq:ModularityDiss1app} or \eqref{eq:ModularityDiss3app} for any parallel computation occurring in finite time, regardless of the local free energies of the memory elements utilized.

\end{widetext}

\section{Deposition and Non-selective Measurement}
\label{sec:NonselectiveMeas}

Eq.~\eqref{eq:EPasWKplusCoherence}
already tells us that any
decoherence on the minimally dissipative eigenbasis
directly contributes to entropy production.
Nevertheless, in the following, we further consider
the loss of superposition and 
the limit of
non-selective measurement, 
which lends a slightly different perspective.

Consider the process of decoherence among a set of decoherence-free subspaces.
To distinguish coherence within these subspaces from coherence \emph{between} the subspaces,
we will refer to the latter as superposition.  The loss of superposition among these subspaces
will be referred to as `deposition', and a state without superposition is a `deposed' state.

In the fully-deposed limit,
this can describe
\emph{non-selective measurement}, which
implements the map
$\rho_0 \mapsto \sum_{m} \Pi_m \rho_0 \Pi_m$,
where the set of projectors $\{ \Pi_m \}_m$
satisfy $\Pi_m \Pi_n = \delta_{m,n} \Pi_m$ and $\sum_{m} \Pi_m = I$.
Notably, these projectors can have arbitrary rank; when the rank is larger than one, the corresponding decoherence-free subspace is nontrivial in the sense of allowing persistent coherence.
More generally, we can consider
partial deposition, where the superposition among these subspaces is reduced but does not need to vanish.

It is profitable to define the set $\Xi$ of deposed states (i.e., non-superposed states):
\begin{align*}
\Xi \equiv \bigl\{ \xi_0 = \sum_m \Pi_m \xi_0 \Pi_m  \bigr\} ~.
\end{align*}
More specifically, these states possess no superposition among the decoherence-free subspaces.
We then consider \emph{deposition operators}---those operations $\Gamma$ that 
cannot create superpositions among the subspaces.  They map deposed states to deposed states:
$\Gamma(\Xi) \subseteq \Xi $.

Let us specifically consider
processes for 
deposition 
that satisfy the following three properties:
\begin{enumerate}
\item 
If the system is already deposed, then the environment does not change.
\item 
Fully deposing the final state yields the same result as evolving the fully deposed state.
\item 
The decoherence-free subspaces evolve unitarily.
\end{enumerate}

More formally, these three properties can be written as:
\begin{align}
\text{If } \rho_0 \in \Xi , \text{ then } \rho_0^\text{env} = \rho_\tau^\text{env} .
\label{eq:DepoProperty1}
\end{align}
\begin{align}
\Pi_m \Gamma(\rho_0) \Pi_m =  \Gamma( \Pi_m \rho_0 \Pi_m ) \label{eq:DepoProperty2}
~.
\end{align}
\begin{align}
\Gamma( \Pi_m \rho_0 \Pi_m ) = U \Pi_m \rho_0 \Pi_m U^\dagger  \;  \text{ for some unitary } U ~. 
\label{eq:DepoProperty3}
\end{align}

The first and third property, Eqs.~\eqref{eq:DepoProperty1} and \eqref{eq:DepoProperty3},
imply, 
via Eqs.~\eqref{eq:EPdef} and \eqref{eq:GenEFdef}, 
that $\EP[\xi_0] = 0$ for $\xi_0 \in \Xi$
since $\Delta S(\xi_t) = 0$ and $\EF = 0$.
By the Second Law, the set of deposed states $\Xi$
are all thus minimally dissipative states for these deposition processes.

For deposition processes with the above three properties, we can  
exactly quantify entropy production from any initial state $\rho_0$
by recognizing $\q_0 = \sum_m \Pi_m \rho_0 \Pi_m \in \Xi$ as a valid choice of the minimally dissipative state 
with $\EP[\q_0] = 0$.
\begin{align}
\EP 
&= \EP - \EP[\q_0] \nonumber \\
&= \! \kB  \text{D} \! \bigl[ \rho_0 \big\| \sum_{m} \Pi_m \rho_0 \Pi_m \bigr] 
    \! - \! \kB  \text{D} \! \bigl[ \rho_\tau \big\| \Gamma \bigl( \sum_{m} \Pi_m \rho_0 \Pi_m \bigr) \bigr]  \nonumber \\
&= - \kB \Delta \text{D} \bigl[ \rho_t \big\| \sum_{m} \Pi_m \rho_t \Pi_m \bigr] ~,
\label{eq:GenDecoherence}
\end{align}
where we have invoked Property \eqref{eq:DepoProperty2} and the linearity of quantum channels to obtain $\Gamma \bigl( \sum_{m} \Pi_m \rho_0 \Pi_m \bigr) = \sum_{m} \Pi_m \Gamma(\rho_0) \Pi_m $.
Eq.~\eqref{eq:GenDecoherence} quantifies entropy production from the loss of superposition 
among decoherence-free subspaces.

In the limit of non-selective measurement, the final state is fully deposed: 
$\rho_\tau = \sum_{m} \Pi_m \rho_\tau \Pi_m$, leading to
\begin{align}
\EP 
= \kB  \text{D} \bigl[ \rho_0 \big\| \sum_{m} \Pi_m \rho_0 \Pi_m \bigr] ~.
\end{align}
This quantifies entropy production when all superposition between subspaces is destroyed.

It is notable that the quantity
$\text{D} \bigl[ \rho_t \big\| \sum_{m} \Pi_m \rho_t \Pi_m \bigr] $---the so-called 
``relative entropy of superposition''~\cite{Aber06}---generalizes the relative entropy of coherence
to allow for decoherence-free subspaces.
When the projection operators are all rank-one (i.e., if $\Pi_m = \ket{m} \bra{m}$ for all $m$), then this quantity reduces to the typical relative entropy of coherence of Ref.~\cite{Baum14}.

\section{Relaxation to NESS}
\label{sec:NESS}

Suppose that the system is in constant contact with at least two different thermodynamic baths.
We may think, for example, of a stovetop pot of water which is hot at its base and cooler at its top surface.
Such a setup famously allows for the existence of nonequilibrium steady states (NESSs),
like Rayleigh--B\'{e}nard convection~\cite{Ahle09}.
Our results---relating entropy production from different initial conditions---should allow interesting new
thermodynamic analyses of such spatiotemporally intricate NESSs.
The thermodynamic behavior could then be compared with the other aspects of
nonequilibrium pattern formation~\cite{Cros93a}.
For Rayleigh--B\'{e}nard convection,
it has been noted that ``the nature of the transient behavior and the eventual
roll locations do depend on the initial state in an unpredictable manner.''~\cite{Rapa88}
In future studies, our results could be leveraged to tie this
phenomenology to thermodynamics.

Another exciting opportunity for future work
would be a more thorough investigation of the relationship between
1) coexisting basins of attraction in the NESS dynamics of nonlinear physical systems
and
2) the thermodynamically independent basins discussed in App.~\ref{sec:Genq0}.
These two `basins' seem to be the same in certain cases, but further careful study will be required to
delineate the general connection
between physical nonlinear dynamics and the implications for its thermodynamics.
	
At a smaller scale,
our results should allow new approaches
to analyzing the thermodynamics of
biomolecules like sodium-ion pumps
or ATP-synthase
that reliably break time symmetry in their NESSs
via differences in chemical potentials
across cellular membranes~\cite{Bust05, Feng08, Seif12}.

While there is not expected to be a general extremization principle for finding
NESSs,
the mere existence of minimally-dissipating initial states---or maximally-dissipating initial states\footnote{Indeed, our main result only depends on the fact that $\q_0$ extremizes $\EP$.}---implies
the in-principle-applicability of our results for
the thermodynamic analyses of general NESSs.
Caveats aside,
there is an obvious opportunity to apply our results to systems with NESSs that \emph{do} extremize entropy production, like certain steady states in the linear regime~\cite{deGr84, Reic09}.

\section{Relation to Error--Dissipation tradeoffs}
\label{sec:TSP}

Under control constraints---like time-symmetric driving---where fidelity costs
significant dissipation~\cite{Riec19a},
we find that $\q_0$ may be forced to have eigenvalues of order $\epsilon$
and thus $\text{D} \bigl[ \rho_0 \big\| \q_0 \bigr] $
can diverge as $\ln(1/\epsilon)$.
This is consistent with the generic error--dissipation tradeoff recently discovered
for non-reciprocated computations in Ref.~\cite{Riec19a},
but only explains the error--dissipation tradeoff for logically irreversible transitions like erasure.

As explained in Ref.~\cite{Riec19a},
\emph{reciprocity} of a memory transition
requires not only logical reversibility, but also a type of logical self-invertibility.
In the case of time-reversal-invariant memory elements,
a deterministic computation $\mathcal{C}$
is reciprocated if its action on a memory state $m$ satisfies
$\mathcal{C}(\mathcal{C}(m)) = m$.
The transition is non-reciprocated otherwise.
For logically reversible but non-reciprocated transitions,
our Theorem 1 and Corollary 1 imply that
all initial distributions could suffer the same
dissipation, since those transitions could be implemented by a unitary transformation that preserves relative entropy.
In those cases of logically reversible non-reciprocity, the error--dissipation tradeoff is \emph{not} a necessary consequence of the contraction of the relative entropy
discussed here, but rather follows more generally from the theory laid out in Ref.~\cite{Riec19a} when time-symmetric control transforms metastable memories.

With unrestricted control, any
arbitrarily-high-fidelity transformation of a finite memory can be achieved with
bounded dissipation.

\section{Generalized derivation for related optimization problems}
\label{sec:RelatedOptimization}

Suppose an initial
product state of the system and environment:
$\rho_0^\text{tot} = \rho_0 \otimes \rho_0^\text{env}$, and suppose that the joint system and environment evolves according to some unitary time evolution, such that the reduced state at time $\tau$ is given by:
$\rho_\tau = \tr_{\text{env}} \bigl( \mathcal{U} \rho_0 \otimes \rho_0^\text{env} \mathcal{U}^\dagger \bigr)$.

We can consider any real-valued functional of the initial density matrix:
\begin{align}
f(\rho_0) = a(\rho_0) + \tr \bigl( \rho_0 \ln \rho_0 - \rho_\tau \ln \rho_\tau \bigr) ~,
\end{align}
and its minimizer:
\begin{align}
\alpha_0 \in \argmin_{\rho_0} f(\rho_0) ~.
\end{align}

Recall Thm.~\ref{thm:GenLinYieldsRelEnt}:
\emph{
If $a(\rho)$ is
an affine
function of $\rho$, 
and
$\alpha_0 \in \argmin_{\rho_0} f(\rho_0)$ has a trivial nullspace,
then
$f(\rho_0) - f(\alpha_0) =
\text{D} \bigl[ \rho_0 \big\| \alpha_0 \bigr] - \text{D} \bigl[ \rho_\tau \big\| \alpha_\tau \bigr] $.}

{\ProThe
If $a(\rho)$ is an affine function, then it can be written as
$a(\rho) = \ell(\rho) + c$, where $\ell(\rho)$ is a linear function of $\rho$ and $c$ is a constant.
Representing
the initial density matrix in an orthonormal basis as
$\rho_0 = \sum_{j, k} c_{j, k} \ket{j} \bra{k}$,
and differentiating $f(\rho_0)$ with respect to the matrix elements of $\rho_0$,
we find
\begin{align}
\frac{\partial}{ \partial c_{j, k} }  f(\rho_0)
&=
\frac{\partial}{ \partial c_{j, k} } \ell(\rho_0)
  + \frac{\partial}{ \partial c_{j, k} } \tr \bigl( \rho_0 \ln \rho_0 - \rho_\tau \ln \rho_\tau \bigr) \\
&=
\ell(\ket{j} \bra{k})
+ \tr \bigl( \ket{j} \bra{k} \ln \rho_0 \bigr)
\nonumber \\
& \quad
 -
\tr \Bigl( \tr_{\text{env}} \bigl(
    \mathcal{U}
    \ket{j} \bra{k}
    \otimes \rho_0^{\text{env}}
    \mathcal{U}^\dagger
   \bigr)
    \ln \rho_\tau
   \Bigr) ~.
\end{align}

To consider the consequences of arbitrary variations in the initial density matrix,
we construct a gradient
$\grad f(\rho_0) \equiv \sum_{j, k} \ket{k} \bra{j} \frac{\partial}{\partial c_{j, k}} f(\rho_0)$
with a scalar product ``$\cdot$'' that gives a type of directional derivative:
$\gamma \cdot \grad f(\rho_0)
\equiv
\tr( \gamma \grad f(\rho_0) )$.

For any two density matrices $\rho_0$ and $\rho_0'$, we find that
\begin{align}
\rho_0 \cdot \! \grad f(\rho_0')
&=
\ell(\rho_0) +
\tr( \rho_0 \ln \rho_0' ) - \tr( \rho_\tau \ln \rho_\tau' )
~.
\label{eq:GenLemma}
\end{align}

Hence,
for any initial density matrix:
$\rho_0 \cdot \grad f(\rho_0)
=
f(\rho_0) - c$.
By definition of
$\alpha_0 \in \argmin_{\rho_0} f(\rho_0)$
as an extremum,
if $\alpha_0$ has full rank,
it must be true that
\begin{align}
(\rho_0 - \alpha_0) \cdot \grad f(\alpha_0) = 0
\label{eq:GenExtremumProperty}
\end{align}
for any density matrix $\rho_0$.
I.e., moving from $\alpha_0$ infinitesimally in the direction of any other initial density matrix cannot
produce a linear change in $f(\rho_0)$.
Expanding Eq.~\eqref{eq:GenExtremumProperty},
$\rho_0 \cdot \grad f(\alpha_0) - \alpha_0 \cdot \grad f(\alpha_0) = 0$,
according to Eq.~\eqref{eq:GenLemma}
yields our generalized result:
\begin{align}
f(\rho_0) - f(\alpha_0) = \text{D} \bigl[ \rho_0 \big\| \alpha_0 \bigr] - \text{D} \bigl[ \rho_\tau \big\| \alpha_\tau \bigr] ~.
\label{eq:GeneralizedResult}
\end{align}
where D$[ \rho \| \alpha ] \equiv \tr( \rho \ln \rho) - \tr(\rho \ln \alpha)$ is the relative entropy.

}

\vspace{1em}

If $\alpha_0$ has a non-trivial nullspace, then Thm.~\ref{thm:GenLinYieldsRelEnt} can be extended
as done in App.~E.

We obtain further interesting results when $a$ is a
nonlinear function---which indicates the growth of other physically relevant quantities (like mutual information with the environment)---and
we will report on these elsewhere.


\begin{thebibliography}{10}

\bibitem{Zure03}
W.~H. Zurek.
\newblock Quantum discord and {M}axwell's demons.
\newblock {\em Phys. Rev. A}, 67:012320, Jan 2003.

\bibitem{Del11}
L.~Del~Rio, J.~{\AA}berg, R.~Renner, O.~Dahlsten, and V.~Vedral.
\newblock The thermodynamic meaning of negative entropy.
\newblock {\em Nature}, 474(7349):61, 2011.

\bibitem{Fais15}
P.~Faist, F.~Dupuis, J.~Oppenheim, and R.~Renner.
\newblock The minimal work cost of information processing.
\newblock {\em Nature communications}, 6:7669, 2015.

\bibitem{Parr15a}
J.~M.~R. Parrondo, J.~M. Horowitz, and T.~Sagawa.
\newblock Thermodynamics of information.
\newblock {\em Nature Physics}, 11(2):131--139, February 2015.

\bibitem{Gool16}
J.~Goold, M.~Huber, A.~Riera, L.~del Rio, and P.~Skrzypczyk.
\newblock The role of quantum information in thermodynamics{\textemdash}a
  topical review.
\newblock {\em Journal of Physics A: Mathematical and Theoretical},
  49(14):143001, feb 2016.

\bibitem{Kolc17}
A.~Kolchinsky and D.~H. Wolpert.
\newblock Dependence of dissipation on the initial distribution over states.
\newblock {\em Journal of Statistical Mechanics: Theory and Experiment},
  2017(8):083202, 2017.

\bibitem{Note1}
More often we must rely on approximations---by assuming weak coupling and
  Markovian dynamics~\cite {Lind76, Alic07}, or leveraging linear response and
  local equilibrium theories~\cite {deGr84}---which have provided practical
  successes in their domain of applicability~\cite {Kubo66, Zwan65, Alic18},
  but cannot be trusted far from equilibrium.

\bibitem{Croo99a}
G.~E. Crooks.
\newblock Entropy production fluctuation theorem and the nonequilibrium work
  relation for free energy differences.
\newblock {\em Phys. Rev. E}, 60:2721, 1999.

\bibitem{Talk07}
P.~Talkner and P.~H\"{a}nggi.
\newblock The {T}asaki{\textendash}{C}rooks quantum fluctuation theorem.
\newblock {\em Journal of Physics A: Mathematical and Theoretical},
  40(26):F569--F571, 2007.

\bibitem{Jarz97a}
C.~Jarzynski.
\newblock Nonequilibrium equality for free energy differences.
\newblock {\em Phys. Rev. Lett.}, 78(14):2690--2693, 1997.

\bibitem{Tasa00}
H.~Tasaki.
\newblock Jarzynski relations for quantum systems and some applications.
\newblock {\em arXiv preprint cond-mat/0009244}, 2000.

\bibitem{Parr09}
J~M~R Parrondo, C~Van den Broeck, and R~Kawai.
\newblock Entropy production and the arrow of time.
\newblock {\em New Journal of Physics}, 11(7):073008, jul 2009.

\bibitem{Mori11}
Y.~Morikuni and H.~Tasaki.
\newblock Quantum {J}arzynski{\textendash}{S}agawa{\textendash}{U}eda
  relations.
\newblock {\em Journal of Statistical Physics}, 143(1):1--10, Apr 2011.

\bibitem{Deff11}
S.~Deffner and E.~Lutz.
\newblock Nonequilibrium entropy production for open quantum systems.
\newblock {\em Physical review letters}, 107(14):140404, 2011.

\bibitem{Aber18}
Johan \AA{}berg.
\newblock Fully quantum fluctuation theorems.
\newblock {\em Phys. Rev. X}, 8:011019, Feb 2018.

\bibitem{Kwon19}
H.~Kwon and M.~S. Kim.
\newblock Fluctuation theorems for a quantum channel.
\newblock {\em Physical Review X}, 9(3):031029, 2019.

\bibitem{Bind18a}
F.~Binder, L.~A. Correa, C.~Gogolin, J.~Anders, and G.~Adesso, editors.
\newblock {\em Thermodynamics in the Quantum Regime: Fundamental Aspects and
  New Directions}.
\newblock Springer, Cham, 2018.

\bibitem{Mica20}
K.~Micadei, G.~T. Landi, and E.~Lutz.
\newblock Quantum fluctuation theorems beyond two-point measurements.
\newblock {\em Phys. Rev. Lett.}, 124:090602, Mar 2020.

\bibitem{Espo10a}
M.~Esposito, K.~Lindenberg, and C.~Van den Broeck.
\newblock Entropy production as correlation between system and reservoir.
\newblock {\em New Journal of Physics}, 12(1):013013, jan 2010.

\bibitem{Reeb14}
D.~Reeb and M.~M. Wolf.
\newblock An improved {L}andauer principle with finite-size corrections.
\newblock {\em New Journal of Physics}, 16(10):103011, 2014.

\bibitem{Dahl17}
O.~C.~O. Dahlsten, M.~Choi, D.~Braun, A.~J.~P. Garner, N.~Y. Halpern, and
  V.~Vedral.
\newblock Entropic equality for worst-case work at any protocol speed.
\newblock {\em New Journal of Physics}, 19(4):043013, 2017.

\bibitem{Halp18}
N.~Y. Halpern, A.~J.~P. Garner, O.~C.~O. Dahlsten, and V.~Vedral.
\newblock Maximum one-shot dissipated work from {R}{\'e}nyi divergences.
\newblock {\em Physical Review E}, 97(5):052135, 2018.

\bibitem{Note2}
While we only utilize the \protect \emph {existence} of the net unitary time
  evolution, we note that it is induced through the time-ordered exponential
  involving the total Hamiltonian $H_{x_t}^\protect \text {tot}$.

\bibitem{deGr84}
S.~R. de~Groot and P.~Mazur.
\newblock {\em Non-equilibrium thermodynamics}.
\newblock Dover Publications, 1984.

\bibitem{Kond14}
D.~Kondepudi and I.~Prigogine.
\newblock {\em Modern thermodynamics: from heat engines to dissipative
  structures}.
\newblock John Wiley \& Sons, 2014.

\bibitem{Reic09}
L.~E. Reichl.
\newblock {\em A Modern Course in Statistical Physics}.
\newblock Wiley-VCH, 2009.

\bibitem{Albe01}
R.~A. Alberty.
\newblock Use of {L}egendre transforms in chemical thermodynamics ({IUPAC}
  technical report).
\newblock {\em Pure and Applied Chemistry}, 73(8):1349--1380, 2001.

\bibitem{Ptas19a}
K.~Ptaszy\ifmmode~\acute{n}\else \'{n}\fi{}ski and M.~Esposito.
\newblock Entropy production in open systems: The predominant role of
  intraenvironment correlations.
\newblock {\em Phys. Rev. Lett.}, 123:200603, Nov 2019.

\bibitem{Hilt11}
S.~Hilt, S.~Shabbir, J.~Anders, and E.~Lutz.
\newblock Landauer’s principle in the quantum regime.
\newblock {\em Physical Review E}, 83(3):030102, 2011.

\bibitem{Jevt12}
S.~Jevtic, D.~Jennings, and T.~Rudolph.
\newblock Maximally and minimally correlated states attainable within a closed
  evolving system.
\newblock {\em Phys. Rev. Lett.}, 108:110403, Mar 2012.

\bibitem{Mica19}
K.~Micadei, J.~P.~S. Peterson, A.~M. Souza, R.~S. Sarthour, I.~S. Oliveira,
  G.~T. Landi, T.~B. Batalh{\~a}o, R.~M. Serra, and E.~Lutz.
\newblock Reversing the direction of heat flow using quantum correlations.
\newblock {\em Nature communications}, 10(1):1--6, 2019.

\bibitem{Timp20}
A.~M. Timpanaro, J.~P. Santos, and G.~T. Landi.
\newblock Landauer's principle at zero temperature.
\newblock {\em Phys. Rev. Lett.}, 124:240601, Jun 2020.

\bibitem{Stin55}
W.~F. Stinespring.
\newblock Positive functions on {C}*-algebras.
\newblock {\em Proceedings of the American Mathematical Society},
  6(2):211--216, 1955.

\bibitem{Hiai91}
F.~Hiai and D.~Petz.
\newblock The proper formula for relative entropy and its asymptotics in
  quantum probability.
\newblock {\em Communications in mathematical physics}, 143(1):99--114, 1991.

\bibitem{Vedr02}
V.~Vedral.
\newblock The role of relative entropy in quantum information theory.
\newblock {\em Reviews of Modern Physics}, 74(1):197, 2002.

\bibitem{Moln10}
L.~Moln{\'a}r and P.~Szokol.
\newblock Maps on states preserving the relative entropy {II}.
\newblock {\em Linear algebra and its applications}, 432(12):3343--3350, 2010.

\bibitem{Buvz99}
V.~Bu{\v{z}}ek, M.~Hillery, and R.~F. Werner.
\newblock Optimal manipulations with qubits: {U}niversal-{NOT} gate.
\newblock {\em Physical Review A}, 60(4):R2626, 1999.

\bibitem{Baum14}
T.~Baumgratz, M.~Cramer, and M.~B. Plenio.
\newblock Quantifying coherence.
\newblock {\em Physical review letters}, 113(14):140401, 2014.

\bibitem{Bran13}
F.~G. S.~L. Brandao, M.~Horodecki, J.~Oppenheim, J.~M. Renes, and R.~W.
  Spekkens.
\newblock Resource theory of quantum states out of thermal equilibrium.
\newblock {\em Physical review letters}, 111(25):250404, 2013.

\bibitem{Aabe14}
J.~{\AA}berg.
\newblock Catalytic coherence.
\newblock {\em Physical review letters}, 113(15):150402, 2014.

\bibitem{Nara19}
V.~Narasimhachar, J.~Thompson, J.~Ma, G.~Gour, and M.~Gu.
\newblock Quantifying memory capacity as a quantum thermodynamic resource.
\newblock {\em Phys. Rev. Lett.}, 122:060601, Feb 2019.

\bibitem{Scan19}
M.~Scandi and M.~Perarnau-Llobet.
\newblock Thermodynamic length in open quantum systems.
\newblock {\em {Quantum}}, 3:197, October 2019.

\bibitem{Note3}
This indeed requires an open quantum system, which introduces the possibility
  of dissipation.

\bibitem{Boyd18b}
A.~B. Boyd, D.~Mandal, and J.~P. Crutchfield.
\newblock Thermodynamics of modularity: structural costs beyond the landauer
  bound.
\newblock {\em Physical Review X}, 8(3):031036, 2018.

\bibitem{Riec19b}
P.~M. Riechers.
\newblock Transforming metastable memories: The nonequilibrium thermodynamics
  of computation.
\newblock In D.~H. Wolpert, C.~Kempes, P.~F. Stadler, and J.~A. Grochow,
  editors, {\em The Energetics of Computing in Life and Machines}, pages
  353--380. SFI Press, 2019.

\bibitem{Loom20a}
S.~P. Loomis and J.~P. Crutchfield.
\newblock Thermodynamically-efficient local computation and the inefficiency of
  quantum memory compression.
\newblock {\em Phys. Rev. Research}, 2:023039, Apr 2020.

\bibitem{Kolc17b}
A.~Kolchinsky, I.~Marvian, C.~Gokler, Z.~Liu, P.~Shor, O.~Shtanko, K.~Thompson,
  D.~Wolpert, and S.~Lloyd.
\newblock Maximizing free energy gain.
\newblock {\em arXiv preprint arXiv:1705.00041}, 2017.

\bibitem{Wolp19}
D.~H. Wolpert and A.~Kolchinsky.
\newblock Thermodynamics of computing with circuits.
\newblock {\em New Journal of Physics}, 22(6):063047, 2020.

\bibitem{Boyd16}
A.~B. Boyd and J.~P. Crutchfield.
\newblock Maxwell demon dynamics: {D}eterministic chaos, the {S}zilard map, and
  the intelligence of thermodynamic systems.
\newblock {\em Physical review letters}, 116(19):190601, 2016.

\bibitem{Boyd15a}
A.~B. Boyd, D.~Mandal, and J.~P. Crutchfield.
\newblock Identifying functional thermodynamics in autonomous {Maxwellian}
  ratchets.
\newblock {\em New J. Physics}, 18:023049, 2016.
\newblock SFI Working Paper 15-07-025; arxiv.org:1507.01537
  [cond-mat.stat-mech].

\bibitem{Aber06}
J.~Aberg.
\newblock Quantifying superposition.
\newblock {\em arXiv preprint quant-ph/0612146}, 2006.

\bibitem{Fran19a}
G.~Francica, J.~Goold, and F.~Plastina.
\newblock Role of coherence in the nonequilibrium thermodynamics of quantum
  systems.
\newblock {\em Phys. Rev. E}, 99:042105, Apr 2019.

\bibitem{Sant19a}
J.~P. Santos, L.~C. C{\'e}leri, G.~T. Landi, and M.~Paternostro.
\newblock The role of quantum coherence in non-equilibrium entropy production.
\newblock {\em npj Quantum Information}, 5(1):23, 2019.

\bibitem{Lena78}
A.~Lenard.
\newblock Thermodynamical proof of the {G}ibbs formula for elementary quantum
  systems.
\newblock {\em Journal of Statistical Physics}, 19(6):575--586, 1978.

\bibitem{Guar18a}
G.~Guarnieri, M.~Kol\'a\ifmmode~\check{r}\else \v{r}\fi{}, and R.~Filip.
\newblock Steady-state coherences by composite system-bath interactions.
\newblock {\em Phys. Rev. Lett.}, 121:070401, Aug 2018.

\bibitem{Gave17}
B.~Gaveau and L.~S. Schulman.
\newblock Decoherence, the density matrix, the thermal state and the classical
  world.
\newblock {\em Journal of Statistical Physics}, 169(5):889--901, 2017.

\bibitem{Gave19}
B.~Gaveau and L.~S. Schulman.
\newblock Decoherence and phase transitions in quantum dynamics.
\newblock {\em Journal of Statistical Physics}, 174(4):800--807, 2019.

\bibitem{Niel10a}
M.~A. Nielsen and I.~L. Chuang.
\newblock {\em Quantum Computation and Quantum Information}.
\newblock Cambridge University Press, Cambridge, United Kingdom, tenth
  anniversary edition, 2010.

\bibitem{Note4}
We adopt the definition used in Refs.~\cite {Aude05, Niel10a}. However, the
  fidelity is sometimes defined as the square of this.

\bibitem{Aude05}
K.~M.~R. Audenaert and J.~Eisert.
\newblock Continuity bounds on the quantum relative entropy.
\newblock {\em Journal of mathematical physics}, 46(10):102104, 2005.

\bibitem{Ahle09}
G.~Ahlers, S.~Grossmann, and D.~Lohse.
\newblock Heat transfer and large scale dynamics in turbulent
  {R}ayleigh{\textendash}{B}{\'e}nard convection.
\newblock {\em Reviews of modern physics}, 81(2):503, 2009.

\bibitem{Cros93a}
M.~C. Cross and P.~C. Hohenberg.
\newblock Pattern formation outside of equilibrium.
\newblock {\em Rev. Mod. Phys.}, 65(3):851--1112, 1993.

\bibitem{Rapa88}
D.~C. Rapaport.
\newblock Molecular-dynamics study of {R}ayleigh{\textendash}{B}{\'e}nard
  convection.
\newblock {\em Physical review letters}, 60(24):2480, 1988.

\bibitem{Bust05}
C.~Bustamante, J.~Liphardt, and F.~Ritort.
\newblock The nonequilibrium thermodynamics of small systems.
\newblock {\em Physics Today}, 58(7):43--48, 2005.

\bibitem{Feng08}
E.~H. Feng and G.~E. Crooks.
\newblock Length of time's arrow.
\newblock {\em Phys. Rev. Lett.}, 101:090602, Aug 2008.

\bibitem{Seif12}
U.~Seifert.
\newblock Stochastic thermodynamics, fluctuation theorems and molecular
  machines.
\newblock {\em Reports on progress in physics}, 75(12):126001, 2012.

\bibitem{Note5}
Indeed, our main result only depends on the fact that $\sigma _0$ extremizes
  $\protect \EP $.

\bibitem{Riec19a}
P.~M. Riechers, A.~B. Boyd, G.~W. Wimsatt, and J.~P. Crutchfield.
\newblock Balancing error and dissipation in computing.
\newblock {\em Phys. Rev. Research}, 2:033524, Sep 2020.

\bibitem{Lind76}
G.~Lindblad.
\newblock On the generators of quantum dynamical semigroups.
\newblock {\em Communications in Mathematical Physics}, 48(2):119--130, 1976.

\bibitem{Alic07}
R.~Alicki and K.~Lendi.
\newblock {\em Quantum dynamical semigroups and applications}, volume 717.
\newblock Springer, 2007.

\bibitem{Kubo66}
R.~Kubo.
\newblock The fluctuation--dissipation theorem.
\newblock {\em Reports on progress in physics}, 29(1):255, 1966.

\bibitem{Zwan65}
R.~Zwanzig.
\newblock Time-correlation functions and transport coefficients in statistical
  mechanics.
\newblock {\em Annual Review of Physical Chemistry}, 16(1):67--102, 1965.

\bibitem{Alic18}
R.~Alicki and R.~Kosloff.
\newblock Introduction to quantum thermodynamics: History and prospects.
\newblock In F.~Binder, L.~A. Correa, C.~Gogolin, J.~Anders, and G.~Adesso,
  editors, {\em Thermodynamics in the Quantum Regime: Fundamental Aspects and
  New Directions}, pages 1--33. Springer, Cham, 2018.


\end{thebibliography}
\end{document}